\documentstyle[12pt,epsfig]{article}
\makeatletter
\def\slash#1{{\mathpalette\c@ncel{#1}}} 
\makeatother

\newcommand\beq{\begin{eqnarray}}
\newcommand\eeq{\end{eqnarray}}

\newcommand\la{\langle}
\newcommand\ra{\rangle}

\def\pslash{\rlap/{\mkern-1mu p}}

\def\sti{\tilde{s}}
\def\tti{\tilde{t}}
\def\uti{\tilde{u}}

\begin{document}
\begin{flushright}
\end{flushright}
\vspace*{15mm}
\begin{center}
{\Large \bf Probing the three-gluon correlation functions
\\[4mm] 
by the single spin asymmetry in $p^\uparrow p\to DX$}
\vspace{1.5cm}\\
 {\sc Yuji Koike$^1$ and Shinsuke Yoshida$^2$}
\\[0.4cm]
\vspace*{0.1cm}{\it $^1$ Department of Physics, Niigata University,
Ikarashi, Niigata 950-2181, Japan}\\
\vspace*{0.1cm}{\it $^2$ Graduate School of Science and Technology, Niigata University,
Ikarashi, Niigata 950-2181, Japan}
\\[3cm]

{\large \bf Abstract} \end{center}
We study the single transverse-spin asymmetry for the inclusive open-charm production
in the $pp$-collision, $p^\uparrow p\to DX$, induced by the three-gluon correlation functions
in the polarized nucleon.  
We derive the corresponding twist-3 cross section formula 
in the leading order with respect to the QCD coupling constant.
As in the case of the semi-inclusive deep inelastic scattering, $ep^\uparrow\to eDX$, 
our result differs from the previous result in the literature.
We also derive a ``master formula" which expresses the 
twist-3 cross section in terms of the $gg\to c\bar{c}$ hard scattering cross section.  
We present a model calculation of the asymmetry at the RHIC energy,
demonstrating the sensitivity of the asymmetry on the form of the three-gluon correlation functions.

\newpage
\section{Introduction}
Open charm production in inclusive hard processes,
such as semi-inclusive deep inelastic scattering (SIDIS), $ep\to eDX$,
and the $D$-meson production in $pp$ collision, $pp\to DX$, is an ideal tool to 
investigate the
gluon distributions in the nucleon.  Similarly the
single spin asymmetry (SSA) in those processes
plays a crucial role to reveal
the multi-gluon correlations in the nucleon\,\cite{Ji92,KQ08,KQVY08,BKTY10,Koike:2010jz,Koike:2011ns}.  
When the transverse momentum 
of the final $D$-meson can be regarded as hard ($P_T\gg \Lambda_{QCD}$),  
one can analyze the processes
in the framework of the collinear factorization\,\cite{ET82,QS92,EKT07}.
In this framework, SSA appears as a twist-3 observable and
can be represented in terms of the multi-parton correlation
functions.  The purely gluonic correlations responsible for SSAs in the open charm
production are represented by  the ``three-gluon" correlation functions.  
\,\footnote{In the framework of the
transverse-momentum-dependent factorization, which is useful to describe the low-$P_T$
hadron production, 
the corresponding gluonic effect is represented as the $k_\perp$-dependent
gluon distribution functions\,\cite{MR01,Anselmino}.}
For the quark-gluon correlation functions in the nucleon, 
there have been many studies in the literature
in connection with SSAs for the light hadron productions\,\cite{ET82}-\cite{Kanazawa:2011bg},
and our understanding on the mechanism of SSA has made a great progress.

In our recent paper\,\cite{BKTY10} we studied the contribution from the three-gluon
correlation functions 
to SIDIS, $ep^\uparrow \to eDX$.  In that study, we have identified the
complete set of the three-gluon correlation functions and 
established the formalism for calculating the twist-3 single-spin-dependent 
cross section induced by those functions.
The gauge invariance and the factorization property of the
cross section have been shown explicitly in the leading order with respect to the
QCD coupling constant.  
The result of that study
differs from the previous study in the literature\,\cite{KQ08}, and 
we clarified the origin of the discrepancy.  
In our another paper\,\cite{Koike:2011ns}, we have developed
a novel ``master formula'' for the three-gluon contribution to $ep^\uparrow\to eDX$
which expresses the corresponding twist-3 cross section in terms of the twist-2 $\gamma^*g\to c\bar{c}$
scattering cross section, extending the similar formula known for the
soft-gluon-pole (SGP) contribution to SSA from the
quark-gluon correlation functions in the nucleon\,\cite{KT071,KT072}.  
This formula simplifies the actual calculation of the twist-3 cross section and
makes its structure transparent, and may be useful to include higher order corrections
to the asymmetry.

The purpose of this paper is to extend these studies to the contribution of the three-gluon
correlation functions to the SSA in the $pp$ collision,
\beq
p^\uparrow(p, S_\perp)+p(p')\to D(P_h)+X, 
\label{ppDX}
\eeq
where $S_\perp$ is the transverse spin vector of the polarized nucleon 
and $P_h$ is the momentum of the $D$-meson satisfying $P_h^2=m_h^2$ with the $D$-meson mass $m_h$. 
The initial nucleons' momenta $p$ and $p'$ are in the collinear configuration and can be regarded as
massless ($p^2=p'^2=0$) in the twist-3 accuracy.    
We will derive a corresponding formula for the twist-3 single-spin-dependent cross section
in the leading order with respect to the QCD coupling constant.  
We will further derive a master formula for (\ref{ppDX}) which connects
the twist-3 cross section to the cross section for the $gg\to c\bar{c}$ scattering.  
We will also present a model calculation for the 
asymmetry $A_N=\Delta\sigma/\sigma\equiv
(\sigma^\uparrow-\sigma^\downarrow)/(\sigma^\uparrow+\sigma^\downarrow)$, 
where $\sigma^{\uparrow(\downarrow)}$ represents the cross section for (\ref{ppDX})
with the initial nucleon polarized along $S_\perp$ ($-S_\perp$), and 
will obtain a constraint on the three-gluon correlation functions, using the
recent RHIC data on $A_N$ for the $D$-meson production\,\cite{Liu}.

The remainder of this paper is organized as follows:
In section 2, we recall the complete set of the three-gluon correlation
functions in the transversely polarized nucleon which are relevant for our study.  
In section 3, we derive the 
twist-2 unpolarized cross section
for the process $pp\to DX$ induced by the gluon-density in the nucleon.  
In section 4, we derive the twist-3 single-spin-dependent cross section 
induced by the three-gluon correlation functions
for the process (\ref{ppDX}), applying the formalism in \cite{BKTY10}.  
In section 5, we derive the master formula 
which expresses the twist-3 cross section for (\ref{ppDX}) induced by the three-gluon
correlation functions in terms of the cross section for $gg\to c\bar{c}$ scattering
in the twist-2 level.  
In section 6, we present a model calculation of $A_N$ for the $D$-meson production
at the RHIC energy, and demonstrate a sensitivity of the asymmetry on the form of the
three-gluon correlation functions.  
Section 7 is devoted to a brief summary.  
In the appendix, we discuss some technical aspects in the derivation of the 
contribution from the initial-state-interaction diagrams to the twist-3 cross section.

\section{Three-gluon correlation functions in the transversely polarized nucleon}

The twist-3 three-gluon correlation functions 
in the transversely polarized nucleon
were first introduced
in \cite{Ji92}.  Then, as was clarified in \cite{BJLO01,Braun09,BKTY10},
there are two independent three-gluon correlation functions
due to the difference in the contraction of color indices.  
Following the notation in \cite{BKTY10}, we call those functions
$O(x_1,x_2)$ and $N(x_1,x_2)$, which are defined from the
lightcone correlation functions of the three field-strengths
of the gluon in the polarized nucleon as
\beq
&&\hspace{-0.8cm}O^{\alpha\beta\gamma}(x_1,x_2)
=-g(i)^3\int{d\lambda\over 2\pi}\int{d\mu\over 2\pi}e^{i\lambda x_1}
e^{i\mu(x_2-x_1)}\la pS|d_{bca}F_b^{\beta n}(0)F_c^{\gamma n}(\mu n)F_a^{\alpha n}(\lambda n)
|pS\ra \nonumber\\
&&=2iM_N\left[
O(x_1,x_2)g^{\alpha\beta}\epsilon^{\gamma pnS_\perp}
+O(x_2,x_2-x_1)g^{\beta\gamma}\epsilon^{\alpha pnS_\perp}
+O(x_1,x_1-x_2)g^{\gamma\alpha}\epsilon^{\beta pnS_\perp}\right]
\label{3gluonO},\\
&&\hspace{-0.8cm}N^{\alpha\beta\gamma}(x_1,x_2)
=-g(i)^3\int{d\lambda\over 2\pi}\int{d\mu\over 2\pi}e^{i\lambda x_1}
e^{i\mu(x_2-x_1)}\la pS|if_{bca}F_b^{\beta n}(0)F_c^{\gamma n}(\mu n)F_a^{\alpha n}(\lambda n)
|pS\ra \nonumber\\
&&=2iM_N\left[
N(x_1,x_2)g^{\alpha\beta}\epsilon^{\gamma pnS_\perp}
-N(x_2,x_2-x_1)g^{\beta\gamma}\epsilon^{\alpha pnS_\perp}
-N(x_1,x_1-x_2)g^{\gamma\alpha}\epsilon^{\beta pnS_\perp}\right].  
\label{3gluonN}
\eeq
where $F_a^{\alpha\beta}\equiv\partial^\alpha A^\beta_a
-\partial^\beta A^\alpha_a +gf_{abc}A_b^\alpha A_c^\beta$ is the gluon's
field strength, and we used the notation $F_a^{\alpha n}\equiv F_a^{\alpha \beta}n_{\beta}$
and $\epsilon^{\alpha pnS_\perp}\equiv \epsilon^{\alpha\mu\nu\lambda}p_\mu n_\nu S_{\perp\lambda}$
with the convention $\epsilon_{0123}=1$.  
$d^{bca}$ and $f^{bca}$ are the symmetric
and anti-symmetric structure constants of the color SU(3) group,
and we have suppressed the gauge-link operators which ensure the gauge invariance.
$p$ is the nucleon momentum, and
$S_\perp$ is the transverse spin vector of the
nucleon normalized as $S_\perp^2=-1$.
In the twist-3 accuracy, $p$ can be regarded as lightlike ($p^2=0$), 
and $n$ is another lightlike vector satisfying $p\cdot n=1$.  To be specific, 
we set $p^\mu=(p^+,0,\mathbf{0}_\perp)$, $n^\mu=(0,n^-, \mathbf{0}_\perp)$, and  
$S^\mu_\perp =(0,0, \mathbf{S}_\perp)$.
The nucleon mass $M_N$ is introduced to define 
$O(x_1,x_2)$ and $N(x_1,x_2)$ dimensionless.  The
decomposition (\ref{3gluonO}) and (\ref{3gluonN})
takes into account all the constraints from hermiticity, 
invariance 
under the parity- and time-reversal transformations and the permutation 
symmetry among the participating three gluon-fields.  The functions
$O(x_1,x_2)$ and $N(x_1,x_2)$ are real and have the following symmetry
properties,
\beq
&&O(x_1,x_2)=O(x_2,x_1),\qquad O(x_1,x_2)=O(-x_1,-x_2),\label{symO}\\
&&N(x_1,x_2)=N(x_2,x_1),\qquad N(x_1,x_2)=-N(-x_1,-x_2).\label{symN}  
\eeq

\section{Twist-2 unpolarized cross section for $pp \to D X$ from gluon-fusion}

We first recall the twist-2 unpolarized cross section for the process (\ref{ppDX})
which is the denominator of $A_N$.  
It receives main contribution from the 
$c\bar{c}$-creation due to the gluon-fusion
with the subsequent fragmentation of the $c$ (or $\bar{c}$) quark
into the $D$ (or $\bar{D}$) meson (Fig. 1).  The 
corresponding unpolarized cross section can be written as 
\beq 
{P_h^0}\frac{d\sigma}{d^3P_h}&=&\frac{\alpha_s^2}
{S}\sum_{f=c,\bar{c}}\int\frac{dx'}{x'}G(x')\int\frac{dz}{z^2}D_f(z)\int\frac{dx}{x}G(x)
\hat{\sigma}_{gg\to c}^U
\delta(\tilde{s}+\tilde{t}+\tilde{u})
\label{unpol}
\eeq
where $S=(p+p')^2$ is the center-of-mass energy squared and $\alpha_s=g^2/(4\pi)$ is the
strong coupling constant.    
The $c\to D$ (or $\bar{c}\to\bar{D}$) fragmentation function $D_f(z)$ 
and the unpolarized gluon distribution in the nucleon $G(x)$ are, respectively, defined as 
\beq
&&\sum_X{1\over N}\int{d\lambda\over 2\pi}e^{-i\lambda/z}\la 0|\psi_i(0)|D(P_h)X\ra
\la D(P_h)X| \bar{\psi}_j(\lambda w)|0\ra
=\left( \pslash_c +m_c\right)_{ij}D_f(z)+\cdots,\\
&&{1\over x}\int{d\lambda\over 2\pi}e^{i\lambda x}\la
p|F_a^{\mu n}(0)F_a^{\nu n}(\lambda n)|p \ra
=-{1\over 2}G(x)g_{\perp}^{\mu\nu}+\cdots, 
\eeq
where $N=3$ is the number of colors, 
$p_c$ is the momentum of the $c$ (or $\bar{c}$) quark fragmenting into the $D$ ($\bar{D}$)-meson 
with $p_c^2=m_c^2$ and 
$w$ is another lightlike vector of $O(1/p^+)$ satisfying $P_{h}\cdot w=1$. 
Thus $p_c$ is related to $P_h$ as $p_c^\mu=P_h^\mu/z+(m_c^2z-m_h^2/z)/2 w^\mu$.  
$g_\perp^{\mu\nu}$ is defined as 
$g_{\perp}^{\mu\nu}\equiv g^{\mu\nu}-p^{\mu}n^{\nu}-p^{\nu}n^{\mu}$. 
The symbol $\cdots$ denotes higher-twist contributions which are irrelevant here. 
The partonic hard cross section $\hat{\sigma}_{gg\to c}^U$
can be obtained from the 9 diagrams shown in Fig. 2
in the leading order (LO) with respect to the QCD coupling constant, and is given by 
\beq
\hat{\sigma}_{gg\to c}^U={1\over 2N}\left({1\over
\tti\uti}-{N\over C_F}{1\over
\sti^2}\right)\left(\tti^2+\uti^2+4m_c^2\sti-{4m_c^4\sti^2\over \tti\uti}\right), 
\label{}
\eeq
where the invariants for $gg\to c\bar{c}$ scattering are defined as 
\beq
\tilde{s}=(xp+x'p')^2\hspace{5mm}\tilde{t}=(p_c-xp)^2-m_c^2\hspace{5mm}\tilde{u}=(p_c-x'p')^2-m_c^2, 
\eeq
and $C_F={N^2-1\over 2N}$.

\begin{figure}[t]
 \begin{center}
  \includegraphics[height=4cm,width=6cm]{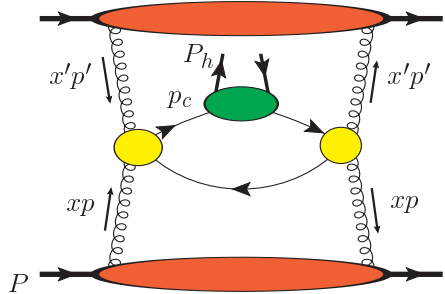}
 \end{center}
 \caption{Generic diagrams for the twist-2 cross section for $pp\to DX$
induced by the gluon densities in the initial nucleons.}
 \label{fig1}
\end{figure}

\begin{figure}[t]
 \begin{center}
  \includegraphics[height=6cm,width=12cm]{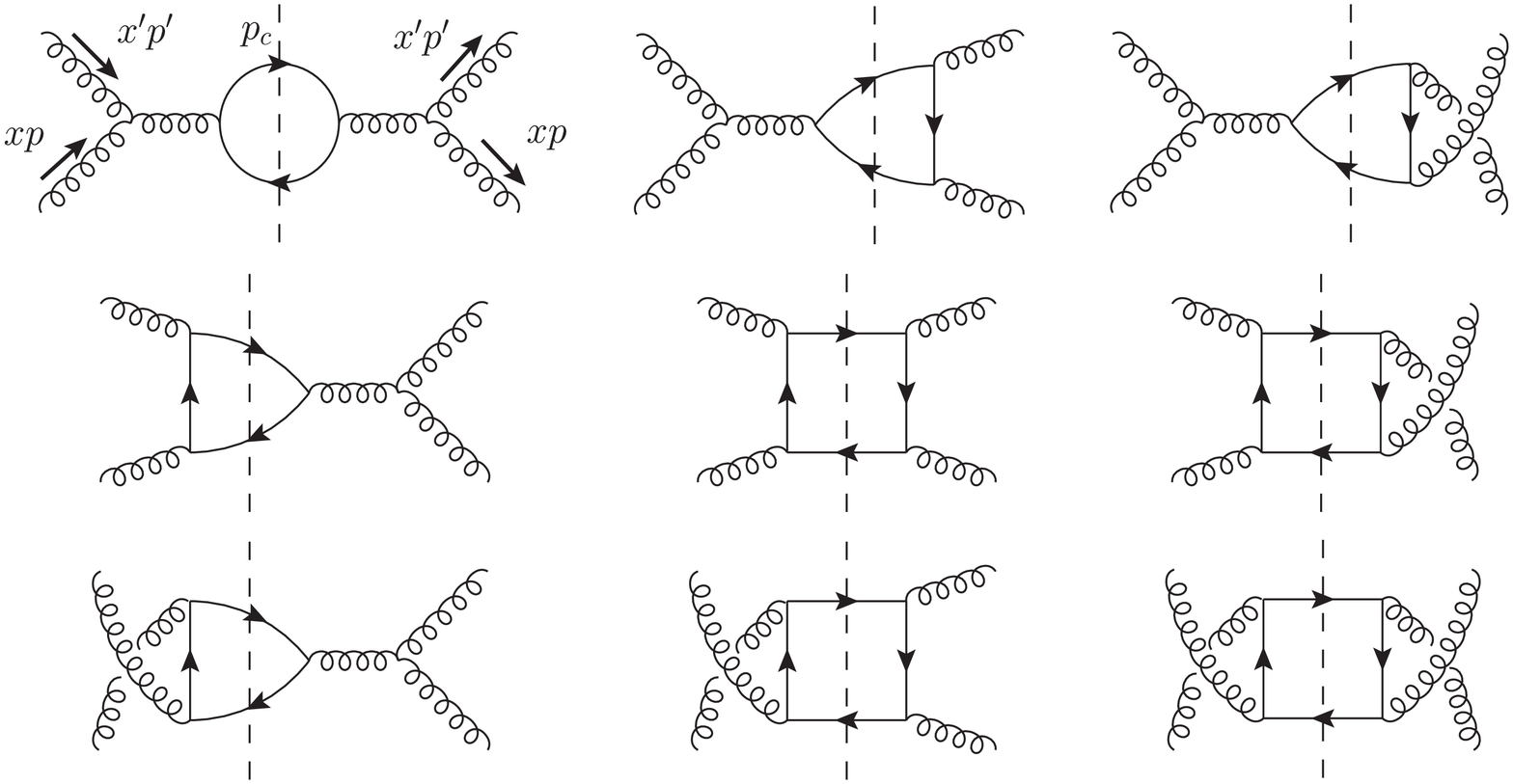}
 \end{center}
 \caption{Leading order diagrams 
for the twist-2 unpolarized hard cross section for $\hat{\sigma}_{gg\to c}^U$
appearing in (\ref{unpol}).}
 \label{fig2}
\end{figure}

\section{Twist-3 cross section for $p^\uparrow p \to D X$ induced by the three-gluon correlation functions}

\begin{figure}[h]
 \begin{center}
  \includegraphics[height=4cm,width=6cm]{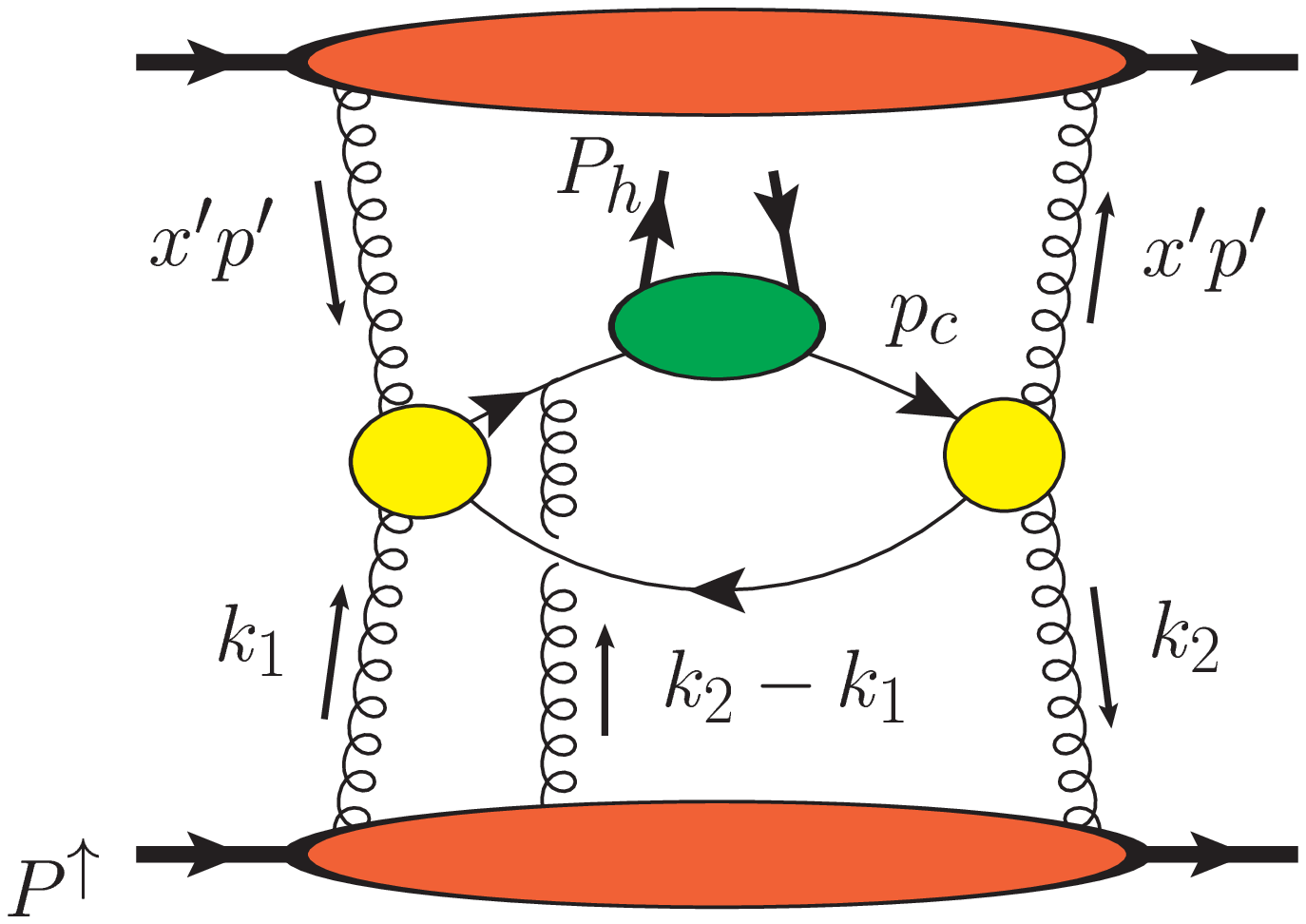}
\qquad\quad
  \includegraphics[height=4cm,width=6cm]{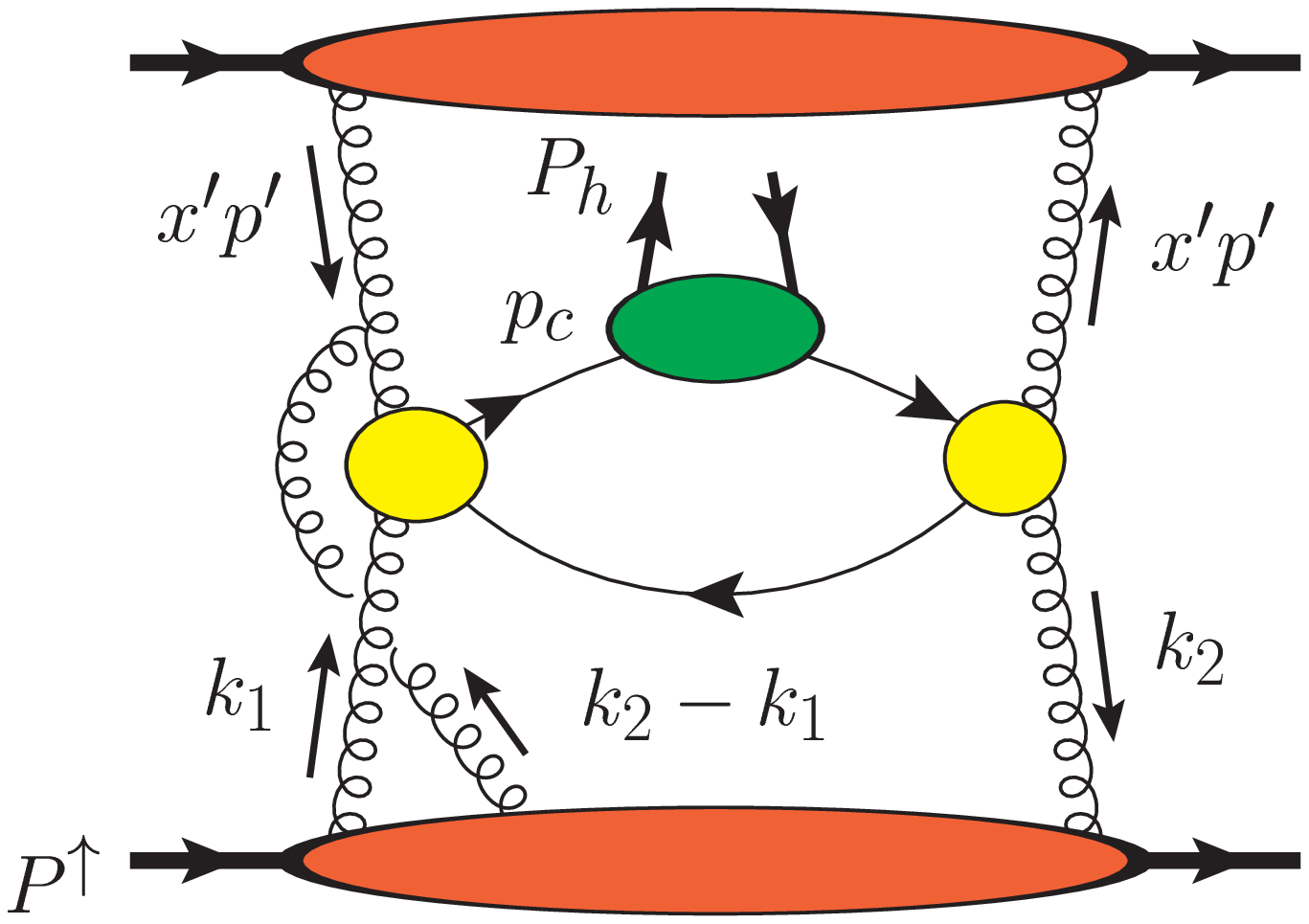}
 \end{center}

\hspace{4.4cm} (a) \hspace{6.5cm} (b) 
 \caption{Generic diagrams which give the twist-3 cross section for $p^\uparrow p\to DX$
induced by the purely gluonic effect in the polarized nucleon (lower blob)
convoluted with the unpolarized gluon density (upper blob) and
the twist-2 fragmentation function for the $D$-meson (middle blob).  A pair of circles in each figure
represent $gg\to c\bar{c}$ hard scattering amplitudes. 
Figures (a) and (b) represent, respectively, the contributions from the final-state-interaction (FSI)
and the initial-state-interaction (ISI). The mirror diagrams also contribute. }
 \label{fig:two}
\end{figure}

The twist-3 single-spin-dependent cross section for $p^\uparrow p \to D X$
induced by the three-gluon correlation functions
can be obtained by applying the formalism developed for
$ep^\uparrow\to eDX$\,\cite{BKTY10}.  
The twist-3 cross section occurs  
from the diagrams of the type shown in Fig. 3, 
where the extra coherent gluon is exchanged between
the hard scattering part and the nucleon matrix element.  
In Fig. 3, the gluon density $G(x')$ in the unpolarized cross section
(upper blob) and the fragmentation function $D_f(z)$ for the $D$-meson (middle blob)
are already factorized.  
Owing to the symmetry property of the
correlation functions (lower blobs of Figs. 3(a) and (b)) defined by
\beq
M_{abc}^{\mu\nu\lambda}(k_1,k_2)=g\int d^4\xi\int d^4\eta e^{ik_1\xi}e^{i(k_2-k_1)\eta}
\la pS | A_b^\nu(0) A_c^\lambda(\eta) A_a^\mu(\xi) |pS\ra,
\label{AAA}
\eeq
the SSA occurs only from a pole part of an internal propagator in the corresponding hard part
\beq
S_{\mu\nu\lambda}^{abc}(k_1,k_2,x'p',p_c), 
\eeq
where
$a$, $b$, $c$ are color indices and the
momenta $k_1$ and $k_2$ are
assigned as shown in Fig. 3.
Here and below we employ the convention that  
the QCD coupling constant $g$ associated with the attachment of the coherent gluon
into the hard part is included in the matrix element (\ref{AAA}) consistently with
the definition of the three-gluon correlation functions in (\ref{3gluonO}) and (\ref{3gluonN}).  
The hard part of the diagrams in Fig. 3 gives rise to the pole contributions at $x_1=x_2$.  
(See discussions below.) 
For those contributions, 
following the same step as \cite{BKTY10} in the collinear expansion to
$S_{\mu\nu\lambda}^{abc}(k_1,k_2,x'p',p_c)$, we eventually end up with
the following expression for the LO
twist-3 cross section induced by the three-gluon correlation function (see Appendix):  
\beq 
P^0_h\frac{d\Delta\sigma}{d^3P_h}&=&\frac{\alpha_s^2}{S}
\sum_{f=c,\bar{c}}\int\frac{dx'}{x'}G(x')\int\frac{dz}{z^2}D_f(z)
\int\frac{dx_1}{x_1}\int\frac{dx_2}{x_2}\nonumber\\
&&\qquad\times
\left[
\left.
{\partial
S_{\mu\nu\lambda}^{abc}(k_1,k_2,x'p',p_c)p^{\lambda}\over 
\partial k_2^{\sigma}}\right|_{k_i=x_ip}
\right]^{\rm pole}
\omega^\mu_{\ \,\alpha}\omega^\nu_{\ \,\beta}\omega^\sigma_{\ \,\gamma}
M^{\alpha\beta\gamma}_{F,abc}(x_1,x_2), 
\label{twist3}
\eeq
where $\omega^\mu_{\ \,\alpha}=g^\mu_{\ \,\alpha}-p^\mu n_\alpha$, and 
$M^{\alpha\beta\gamma}_{F,abc}(x_1,x_2)$ is the
lightcone correlation function of the field-strengths defined as 
\beq
\hspace{-0.5cm}
M^{\alpha\beta\gamma}_{F,abc}(x_1,x_2)
&=&-g(i)^3\int{d\lambda\over 2\pi}\int{d\mu\over 2\pi}e^{i\lambda x_1}
e^{i\mu(x_2-x_1)}\la pS|F_b^{\beta n}(0)F_c^{\gamma n}(\mu n)F_a^{\alpha n}(\lambda n)
|pS\ra \nonumber\\
&=&{N d_{bca}\over (N^2-4)(N^2-1)}
O^{\alpha\beta\gamma}(x_1,x_2)-{if_{bca}\over N(N^2-1)}N^{\alpha\beta\gamma}(x_1,x_2)
\label{Ffunction}
\eeq
with $O^{\alpha\beta\gamma}(x_1,x_2)$ and $N^{\alpha\beta\gamma}(x_1,x_2)$ 
defined in (\ref{3gluonO}) and (\ref{3gluonN}), respectively. 
The symbol $[\cdots]^{\rm pole}$ indicates the pole contribution is to be taken from the hard part.  
We emphasize that even though the analysis of Fig. 3 starts with
the gauge-noninvariant correlation function (\ref{AAA})
and the corresponding hard part $S_{\mu\nu\lambda}^{abc}(k_1,k_2,x'p',p_c)$, 
gauge-noninvariant contributions appearing in the collinear expansion
either vanish or cancel and 
the total surviving twist-3 contribution to the
single-spin-dependent cross section can be expressed as in (\ref{twist3}), using the
gauge-invariant correlation functions (\ref{3gluonO}) and (\ref{3gluonN}).

\begin{figure}[h]
 \begin{center}
  \includegraphics[height=5.2cm,width=11.5cm]{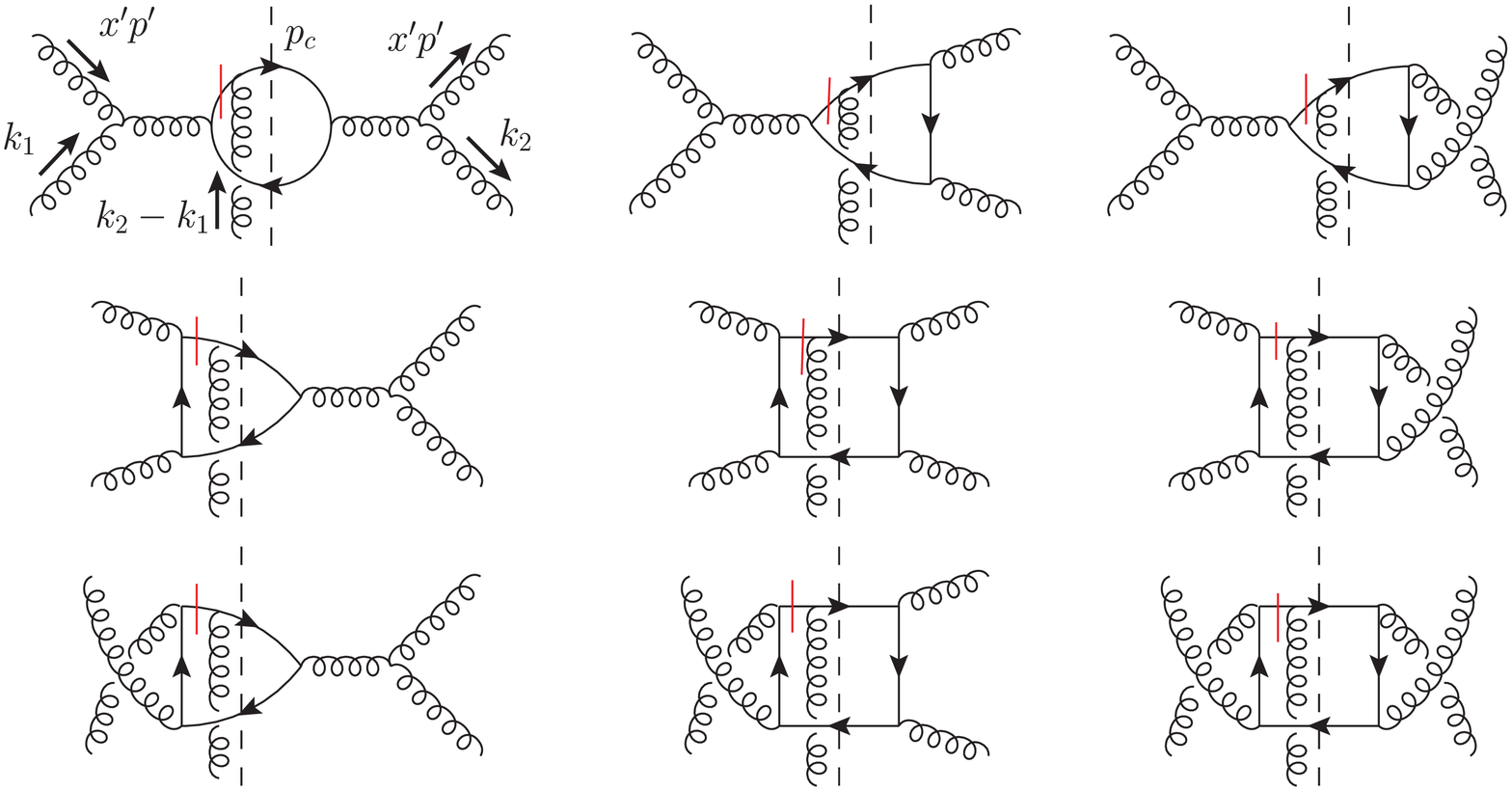}
\vspace{0.2cm}

(a)

\vspace{0.5cm}
  \includegraphics[height=5.2cm,width=11.5cm]{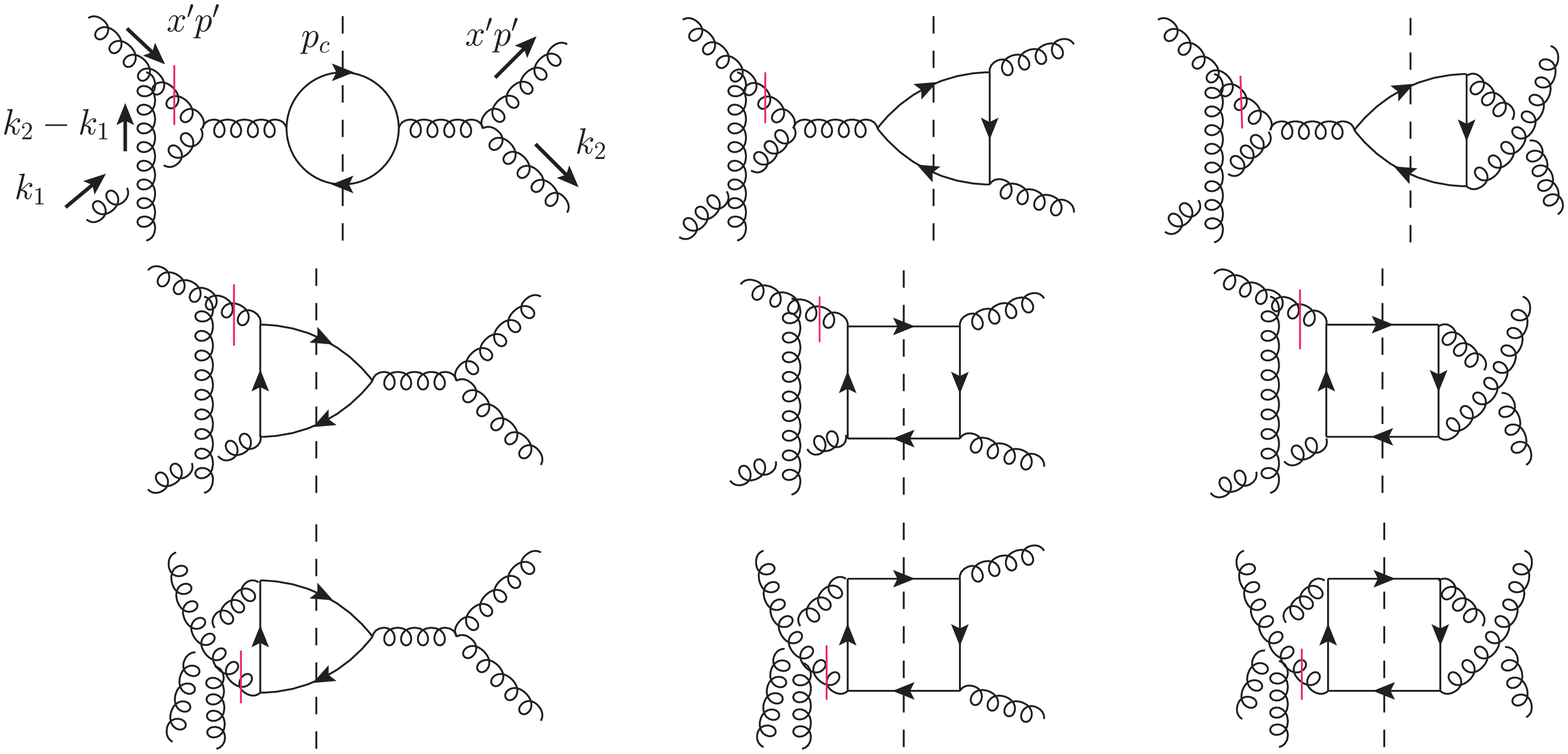}
\vspace{0.2cm}  

(b)
 \end{center}
 \caption{The LO diagrams for the partonic hard part 
$S_{\mu\nu\lambda}^{abc}(k_1,k_2,x'p',p_c)$ for the twist-3 cross section.
Diagrams in (a) represent the FSI contribution and those in (b) represent the ISI contribution.}
 \label{fig:two}
\end{figure}

The pole contribution to the hard part $\left[\left.
{\partial
S_{\mu\nu\lambda}^{abc}(k_1,k_2,x'p',p_c)p^{\lambda} /
\partial k_2^{\sigma}}\right|_{k_i=x_ip}\right]^{\rm pole}$
occurs from 
two types of diagrams shown in Figs. 3 (a) and 3(b), which are refered to as 
the final state interaction (FSI) and the initial state interaction (ISI), respectively,
by the parton lines to which the coherent gluon is attached.   
Figs. 4 (a) and 4(b) show the LO diagrams contributing 
to the hard part $S_{\mu\nu\lambda}^{abc}(k_1,k_2,x'p',p_c)$ in Figs. 3(a) (FSI) and 3(b) (ISI), respectively.    
The mirror diagrams of Fig. 4 also contribute.
The poles are produced from the bared propagator, and gives rise to the
$\delta$-function at $x_1=x_2$ in the collinear limit ($k_i\to x_ip$), hence the poles are
refered to
as the soft-gluon-pole (SGP).  
Other pole contributions cancel among each other by taking the sum of the whole diagrams.

By calculating 
 $\left[\left.
{\partial
S_{\mu\nu\lambda}^{abc}(k_1,k_2,x'p',p_c)p^{\lambda}/
\partial k_2^{\sigma}}\right|_{k_i=x_ip}\right]^{\rm pole}$ from Fig. 4
contracted with
the coefficient tensors in the decomposition of (\ref{3gluonO}) and (\ref{3gluonN}), one obtains 
the twist-3 single-spin-dependent cross section as\,\cite{Koike:2010jz}
\small
 \begin{eqnarray}
&&\hspace{-0.3cm}
P_h^0\frac{d\Delta\sigma}{d^3P_h}=\frac{\alpha_s^2M_N\pi}{S}\epsilon^{P_h p n S_{\perp}}
\sum_{f=c\bar{c}}\int\frac{dx'}{x'}G(x')\int\frac{dz}{z^2}D_f(z)\int\frac{dx}{x}\delta
 \left(\tilde{s}+\tilde{t}+\tilde{u}\right){1\over z\tilde{u}}
 \nonumber\\
&&\hspace{-0.3cm}
\times\biggl[\delta_f\left\{
\left(\frac{d}{dx}O(x,x)-\frac{2O(x,x)}{x}\right)\hat{\sigma}^{O1}
+\left(\frac{d}{dx}O(x,0)-\frac{2O(x,0)}{x}\right)\hat{\sigma}^{O2}
+\frac{O(x,x)}{x}\hat{\sigma}^{O3}
+\frac{O(x,0)}{x}\hat{\sigma}^{O4}
\right\} \nonumber\\
&&\hspace{-0.3cm}
+\left\{
\left(\frac{d}{dx}N(x,x)-\frac{2N(x,x)}{x}\right)\hat{\sigma}^{N1}
+\left(\frac{d}{dx}N(x,0)-\frac{2N(x,0)}{x}\right)\hat{\sigma}^{N2}
+\frac{N(x,x)}{x}\hat{\sigma}^{N3}
+\frac{N(x,0)}{x}\hat{\sigma}^{N4}
\right\}
\biggr].\nonumber\\
\label{twist3final}
\end{eqnarray}
\normalsize
where 
$\delta_c=1$ and 
$\delta_{\bar{c}}=-1$.  The partonic hard cross sections are given by 
\begin{eqnarray}
\left\{
\begin{array}{lll}
\hat{\sigma}^{O1}&=&\displaystyle
\left(\frac{1}{C_F}\frac{\tilde{u}-\tilde{t}}{\tilde{s}\tilde{t}\tilde{u}}
+\frac{1}{C_F}\frac{\tilde{u}}{\tilde{s}\tilde{t}^2}
-\frac{1}{N^2C_F}\frac{\tilde{s}}{\tilde{t}^2\tilde{u}}\right)\left(\tilde{t}^2+\tilde{u}^2+4m_c^2\tilde{s}
-\frac{4m_c^4\tilde{s}^2}{\tilde{t}\tilde{u}}\right),\\[12pt]
 \hat{\sigma}^{O2}&=&\displaystyle
\left(\frac{1}{C_F}\frac{\tilde{u}-\tilde{t}}{\tilde{s}\tilde{t}\tilde{u}}+\frac{1}{C_F}\frac{\tilde{u}}{\tilde{s}\tilde{t}^2}-\frac{1}{N^2C_F}\frac{\tilde{s}}{\tilde{t}^2\tilde{u}}\right)\left(\tilde{t}^2+\tilde{u}^2+8m_c^2\tilde{s}-\frac{8m_c^4\tilde{s}^2}{\tilde{t}\tilde{u}}\right), \\[12pt]
 \hat{\sigma}^{O3}&=&\displaystyle\left(\frac{1}{C_F}\frac{\tilde{u}-\tilde{t}}{\tilde{t}^2\tilde{u}^2}+\frac{1}{C_F}\frac{1}{\tilde{t}^3}-\frac{1}{N^2C_F}\frac{\tilde{s}^2}{\tilde{t}^3\tilde{u}^2}\right)\left(8m_c^4\tilde{s}-4m_c^2\tilde{t}\tilde{u}\right),\\[12pt]
 \hat{\sigma}^{O4}&=&\displaystyle\left(\frac{1}{C_F}\frac{\tilde{u}-\tilde{t}}{\tilde{t}^2\tilde{u}^2}+\frac{1}{C_F}\frac{1}{\tilde{t}^3}-\frac{1}{N^2C_F}\frac{\tilde{s}^2}{\tilde{t}^3\tilde{u}^2}\right)\left(16m_c^4\tilde{s}-4m_c^2\tilde{t}\tilde{u}\right), \\
\end{array}
\right.
\label{hardO}
\end{eqnarray}
and
\begin{eqnarray}
\left\{
\begin{array}{lll}
 \hat{\sigma}^{N1}&=&\displaystyle\left(\frac{1}{C_F}\frac{\tilde{t}^2
+\tilde{u}^2}{\tilde{s}^2\tilde{t}\tilde{u}}+\frac{1}{C_F}\frac{\tilde{u}}{\tilde{s}\tilde{t}^2}-\frac{1}{N^2C_F}\frac{\tilde{s}}{\tilde{t}^2\tilde{u}}\right)\left(\tilde{t}^2+\tilde{u}^2+4m_c^2\tilde{s}-\frac{4m_c^4\tilde{s}^2}{\tilde{t}\tilde{u}}\right), \\[12pt]
 \hat{\sigma}^{N2}&=&\displaystyle-\left(\frac{1}{C_F}\frac{\tilde{t}^2+\tilde{u}^2}{\tilde{s}^2\tilde{t}\tilde{u}}+\frac{1}{C_F}\frac{\tilde{u}}{\tilde{s}\tilde{t}^2}
-\frac{1}{N^2C_F}\frac{\tilde{s}}{\tilde{t}^2\tilde{u}}\right)\left(\tilde{t}^2+\tilde{u}^2+8m_c^2\tilde{s}-
\frac{8m_c^4\tilde{s}^2}{\tilde{t}\tilde{u}}\right), \\[12pt]
 \hat{\sigma}^{N3}&=&\displaystyle\left(\frac{1}{C_F}\frac{\tilde{t}^2+\tilde{u}^2}{\tilde{s}\tilde{t}^2\tilde{u}^2}+\frac{1}{C_F}\frac{1}{\tilde{t}^3}-\frac{1}{N^2C_F}\frac{\tilde{s}^2}{\tilde{t}^3\tilde{u}^2}\right)\left(8m_c^4\tilde{s}-4m_c^2\tilde{t}\tilde{u}\right), \\[12pt]
 \hat{\sigma}^{N4}&=&-\displaystyle\left(\frac{1}{C_F}\frac{\tilde{t}^2
+\tilde{u}^2}{\tilde{s}\tilde{t}^2\tilde{u}^2}+\frac{1}{C_F}\frac{1}{\tilde{t}^3}-\frac{1}{N^2C_F}\frac{\tilde{s}^2}{\tilde{t}^3\tilde{u}^2}\right)\left(16m_c^4\tilde{s}-4m_c^2\tilde{t}\tilde{u}\right), \\
\end{array}
\right.
\label{hardN}
\end{eqnarray}
The hard cross sections associated with the first term in the first parentheses in
(\ref{hardO}) and (\ref{hardN}) come from ISI (Fig. 4(b)), and those associated
with the second and the third terms in the same parentheses come from FSI (Fig. 4(a)).  
As in the case of $ep^\uparrow\to eDX$, the cross section in (\ref{twist3final}) 
receives the contribution from the
four functions $O(x,x)$, $O(x,0)$, $N(x,x)$ and $N(x,0)$. 
Unlike the case of SIDIS, presence of ISI gives rise to the different hard cross sections
for $O$ and $N$ functions.  
From (\ref{twist3final}), it is clear that the process $p^\uparrow p\to DX$ itself 
is not sufficient for the complete separation of the four functions.
For the separation, the process $ep^\uparrow\to eDX$ serves greatly, since it has five
structure functions with different dependences on the azimuthal angles to which 
the four functions contribute differently\,\cite{BKTY10}.  
For the massless quark fragmenting into a light hadron (i.e. $m_c\to 0$ ), one 
has $\hat{\sigma}^{O3,O4,N3,N4}\to 0$, $\sigma^{O1}=\sigma^{O2}$ and $\sigma^{N1}=-\sigma^{N2}$.  
Therefore the three-gluon correlation functions appear in the combination
of $x(d/dx)(O(x,x)+O(x,0))-2(O(x,x)+O(x,0))$ and
$x(d/dx)(N(x,x)-N(x,0))-2(N(x,x)-N(x,0))$ at $m_c=0$.

Our result in (\ref{twist3final}) differs from a previous work\,\cite{KQVY08}:  
The result in \cite{KQVY08} is obtained from (\ref{twist3final}) by omitting the terms
with $\hat{\sigma}^{O2,O4}$ and $\hat{\sigma}^{N2,N4}$
and by the replacement $O(x,x)\to O(x,x)+O(x,0)$ and $N(x,x)\to N(x,x)-N(x,0)$.  
This difference originates from an ad-hoc assumption in the factorization formula
in \cite{KQVY08,KQ08}.  We emphasize the appearance of the four different contributions with   
$\{O(x,x), O(x,0), N(x,x), N(x,0)\}$ is a consequence of the symmetry property
implied in the decomposition (\ref{3gluonO}) and (\ref{3gluonN}), in particular, 
the different coefficient tensors in front of $O(x,x)$ and $O(x,0)$ (likewise for $N(x,x)$ and $N(x,0)$)
at $x_1=x_2=x$ 
lead to different hard cross sections for the above four functions.   
See \cite{BKTY10} for more details.

\section{Master formula for the three-gluon contribution to $p^\uparrow p\to DX$}
\subsection{Connection between the twist-3 cross section and the $gg\to c\bar{c}$ scattering}

To obtain the twist-3 cross section based on (\ref{twist3}), one has to calculate
the derivative of the hard part
$[\left.
{\partial
S_{\mu\nu\lambda}^{abc}(k_1,k_2,x'p',p_c)p^{\lambda}/
\partial k_2^{\sigma}}\right|_{k_i=x_ip}]^{\rm pole}$ 
from Fig. 4 
contracted with
the coefficient tensors in the decomposition of (\ref{3gluonO}) and (\ref{3gluonN}).
This calculation 
produces lots of terms at the intermediate step and is extremely complicated.  
Alternatively, 
application of the ``master formula" developed for the contribution of 
the quark-gluon correlation functions\,\cite{KT071,KT072} 
and also for the three-gluon correlations for $ep^\uparrow\to eDX$\,\cite{Koike:2011ns}
provides us with a more transparent and simpler method to
calculate the cross section.  To extend the method to the 
contribution of the three-gluon correlation
functions for $p^\uparrow p\to DX$, we first note that the diagrams 
in Fig. 4 are
obtained by attaching the extra gluon-line 
to the external-lines of the twist-2 hard part in Fig. 2, and the pole contribution
is given by the propagator next to the vertex to which this extra-gluon line attaches.   
Because of this structure, the derivative can be performed by keeping the structure
of the hard part corresponding to those in Fig. 2 almost intact. 
Based on this observation, one can obtain the master formula also for the three-gluon contribution. 
To be specific, we consider the case in which the $c$-quark fragments into the $D$-meson below.   

To present the result, we first define the hard part for the unpolarized cross section shown in Fig. 2
as ${\cal H}_{\mu\nu}^{U,ab}(xp,x'p',p_c)$, where $\mu\nu$ and $ab$ are,
respectively, the Lorentz and the color indices
for the gluon line with the momentum $xp$.  Those
indices for the gluon line with the momentum $x'p'$ are already contracted
to factorize $G(x')$ in (\ref{unpol}).  With this convention
the partonic hard cross section $\hat{\sigma}_{gg\to c}$ in (\ref{unpol})
is related to ${\cal H}_{\mu\nu}^{U,ab}$ as
\beq
\hat{\sigma}_{gg\to c}^U(\tilde{s},\tilde{t},\tilde{u},m_c^2) 
\delta\left(\tilde{s}+\tilde{t}+\tilde{u}\right)
={1\over (N^2-1)}\delta_{ab}\left(-{1\over 2}g_{\perp}^{\mu\nu}\right)
{\cal H}^{U,ab}_{\mu\nu}(xp,x'p',p_c).
\label{unpolH}
\eeq
As is shown below the hard part for the twist-3 cross section 
has a simple relation with this ${\cal H}^{U,ab}_{\mu\nu}(xp,x'p',p_c)$. 

In \cite{Koike:2011ns}, we have shown that the twist-3 hard cross section for $ep^\uparrow\to eDX$
induced by the three-gluon correlation functions can be obtained
from the Born cross section for the $\gamma^* g\to c\bar{c}$ scattering.  
There the SGP contribution occurs from the FSI.  
Accordingly, 
the FSI contribution for $p^\uparrow p\to DX$ shown in Fig. 4(a) can also be expressed
in terms of the Born cross section for the $g g\to c\bar{c}$ scattering.  
We write the FSI contribution to $S_{\mu\nu\lambda}^{abc}(k_1,k_2,x'p',p_c)p^{\lambda}$
in (\ref{twist3})
as $S_{\mu\nu\lambda}^{F,abc}(k_1,k_2,x'p',p_c)p^{\lambda}$.  
Then, following the same procedure as \cite{Koike:2011ns}, one can show that the hard part for the 
FSI is given by
\beq
&&\left[
\left.
{\partial
S_{\mu\nu\lambda}^{F,abc}(k_1,k_2,x'p',p_c)p^{\lambda}\over 
\partial k_2^{\sigma}}\right|_{k_i=x_ip}\right]^{\rm pole}\nonumber\\
&&\qquad= \left[{1\over x_1 - x_2 +i\epsilon}\right]^{\rm pole} 
\left( {\partial \over \partial p_c^\sigma}
-{p_{c\sigma} p^\lambda \over p\cdot p_c} {\partial \over \partial p_c^\lambda} \right) 
{\cal H}^{F,abc}_{\mu\nu}(x_1p,x'p',p_c)\nonumber\\
&&\qquad= \left[{1\over x_1 - x_2 +i\epsilon}\right]^{\rm pole}
{d\over d p_c^\sigma}{\cal H}^{F,abc}_{\mu\nu}(x_1p,x'p',p_c),
\label{masterF}
\eeq
where ${\cal H}^{F,abc}_{\mu\nu}(x_1p,x'p',p_c)$ is obtained 
from ${\cal H}^{U,ab}_{\mu\nu}(x_1p,x'p',p_c)$
simply by adding the extra color matrix $t^c$ in the same place where
the coherent gluon line is attached in Fig. 4(a).  
Here one needs to be cautious in  
taking derivative with respect to $p_c^\sigma$:  
In the expression after the first equality of (\ref{masterF}), the on-shell limit $p_c^2=m_c^2$ 
should be taken after performing the derivative with respect to $p_c^\sigma$.    
For the derivative in the expression after the second equality of (\ref{masterF}), 
the form
$p_c^\mu=\left( p^+_c={m_c^2 + \vec{p}_{c\perp}^{\,2} \over 2p^-}, p^-_c, \vec{p}_{c\perp}\right)$
should be used for $p_c$, i.e., on-shell condition for $p_c$ should be used
by regarding $p_c^+$ as a dependent variable of $p_c^-$ and $\vec{p}_{c\perp}$.  

One can also derive the similar relation for the ISI contribution.  
We write the ISI contribution to $S_{\mu\nu\lambda}^{abc}(k_1,k_2,x'p',p_c)p^{\lambda}$
in (\ref{twist3})
as $S_{\mu\nu\lambda}^{I,abc}(k_1,k_2,x'p',p_c)p^{\lambda}$.  
For the ISI diagrams in Fig. 4(b),   
the coherent gluon couples to the initial gluon-line of the diagrams in Fig. 2 through the three-gluon
coupling.   One can still
apply the same method as \cite{Koike:2011ns}, and obtains for the ISI contribution as
\beq
&&\left[\left.{\partial S_{\mu\nu\lambda}^{I,abc}(k_1,k_2,x'p',p_c)p^{\lambda}\over 
\partial k_2^{\sigma}}\right|_{k_i=x_ip}\right]^{\rm pole} \nonumber\\
&&\qquad = \left[{-1\over x_2 - x_1 +i\epsilon}\right]^{\rm pole} 
 \left( {\partial \over \partial (x'p'^\sigma)}
-{p'_{\sigma} p^\lambda \over p\cdot p'} {\partial \over \partial (x'p'^\lambda)} \right) 
{\cal H}^{I,abc}_{\mu\nu}(x_1p,x'p',p_c),\nonumber\\
&&\qquad = \left[{-1\over x_2 - x_1 +i\epsilon}\right]^{\rm pole} 
{d\over d (x'p'^\sigma)}{\cal H}^{I,abc}_{\mu\nu}(x_1p,x'p',p_c), 
\label{masterI}
\eeq
where ${\cal H}^{I,abc}_{\mu\nu}(x_1p,x'p',p_c)$ differs
from ${\cal H}^{U,ab}_{\mu\nu}(x_1p,x'p',p_c)$ only with its
extra color index $c$ associated with the attachment of the coherent gluon line in Fig. 4(b).  
As in (\ref{masterF}) the on-shell limit $p'^2\to 0$ 
should be taken after carrying out 
the derivative in the expression after the first equality in (\ref{masterI}), and
the on-shell form 
$p'^\mu=\left( p'^+={ \vec{p'}_{\perp}^{\,2} \over 2p'^-}, p'^-, \vec{p'}_{\perp}\right)$
should be used
in the expression after the second equality of (\ref{masterI}).  
In (\ref{masterI}), we first make ${p'}_\perp^\sigma\neq 0$ in taking the derivative and then
take the ${p'}_\perp^\sigma\to 0$ limit to consider the cross section in the frame where $p$ and $p'$
are collinear.    
Inserting (\ref{masterF}) and (\ref{masterI}) into (\ref{twist3}), one obtains
the single-spin-dependent cross section as
\beq 
P_h^0\frac{d\Delta\sigma}{d^3P_h}&=&\frac{\alpha_s^2}{S}
\int\frac{dx'}{x'}G(x')\int\frac{dz}{z^2}D_c(z)\int\frac{dx}{x^2}(-i\pi)
\omega^\mu_{\ \,\alpha}\omega^\nu_{\ \,\beta}\omega^\sigma_{\ \,\gamma}
M^{\alpha\beta\gamma}_{F,abc}(x,x)
\nonumber\\
&&\qquad\times
\left[
{d\over dp_c^\sigma}{\cal H}^{F,abc}_{\mu\nu}(xp,x'p',p_c) -{d\over d(x'p'^\sigma)}
{\cal H}^{I,abc}_{\mu\nu}(xp,x'p',p_c)\right].  
\label{master}
\eeq
The hard part
${\cal H}^{F,abc}_{\mu\nu}$ and ${\cal H}^{I,abc}_{\mu\nu}$ 
contain the factor $\delta\left((xp+x'p'-p_c)^2-m_c^2\right)
=\delta\left(\tilde{s}+\tilde{t}+\tilde{u}\right)$
as an on-shell condition for the final unobserved $\bar{c}$-quark.    
For convenience we separate this $\delta$-function and introduce the 
following functions by taking the color contraction:  
\beq
&&{Nd_{bca}\over (N^2-1)(N^2-4)}
{\cal H}_{\alpha\beta}^{F,abc}(xp,x'p',p_c)
\equiv H^{(F,d)}_{\alpha\beta}(xp,x'p',p_c)
\delta\left(\tilde{s}+\tilde{t}+\tilde{u}\right),\nonumber\\
&&{-if_{bca}\over N(N^2-1)}
{\cal H}_{\alpha\beta}^{F,abc}(xp,x'p',p_c)
\equiv H^{(F,f)}_{\alpha\beta}(xp,x'p',p_c)
\delta\left(\tilde{s}+\tilde{t}+\tilde{u}\right),\nonumber\\
&&{Nd_{bca}\over (N^2-1)(N^2-4)}
{\cal H}_{\alpha\beta}^{I,abc}(xp,x'p',p_c)
\equiv H^{(I,d)}_{\alpha\beta}(xp,x'p',p_c)
\delta\left(\tilde{s}+\tilde{t}+\tilde{u}\right),\nonumber\\
&&{-if_{bca}\over N(N^2-1)}
{\cal H}_{\alpha\beta}^{I,abc}(xp,x'p',p_c)
\equiv H^{(I,f)}_{\alpha\beta}(xp,x'p',p_c)
\delta\left(\tilde{s}+\tilde{t}+\tilde{u}\right).
\label{Hfunction}
\eeq
Using these forms in (\ref{master}), one can write the cross section as 
\beq 
&&\hspace{-0.7cm}P_h^0\frac{d\Delta\sigma}{d^3P_h}=\frac{\alpha_s^2}{S}
\int\frac{dx'}{x'}G(x')\int\frac{dz}{z^2}D_c(z)\int\frac{dx}{x^2}(-i\pi)
\nonumber\\
&&\times\left[
O_\perp^{\alpha\beta\gamma}(x,x)\left\{
{d\over dp_c^\gamma}{H}^{(F,d)}_{\alpha\beta}(xp,x'p',p_c) 
-{d\over d(x'p'^\gamma)}{H}^{(I,d)}_{\alpha\beta}(xp,x'p',p_c)
\right\}\delta\left(\tilde{s}+\tilde{t}+\tilde{u}\right)\right.\nonumber\\
&&\left.+N_\perp^{\alpha\beta\gamma}(x,x)\left\{
{d\over dp_c^\gamma}{H}^{(F,f)}_{\alpha\beta}(xp,x'p',p_c) 
-{d\over d(x'p'^\gamma)}{H}^{(I,f)}_{\alpha\beta}(xp,x'p',p_c)
\right\}\delta\left(\tilde{s}+\tilde{t}+\tilde{u}\right)
\right],
\label{master2}
\eeq
where $O_\perp^{\alpha\beta\gamma}(x,x)$ and 
$N_\perp^{\alpha\beta\gamma}(x,x)$ are the functions obtained by setting $x_1=x_2=x$ in 
(\ref{3gluonO}) and (\ref{3gluonN}):  
\beq
&&O_\perp^{\alpha\beta\gamma}(x,x)=2iM_N\left[
O(x,x)g_\perp^{\alpha\beta}\epsilon^{\gamma pnS_\perp}
+O(x,0)(g_\perp^{\beta\gamma}\epsilon^{\alpha pnS_\perp}
+g_\perp^{\gamma\alpha}\epsilon^{\beta pnS_\perp})\right],\nonumber\\
&&N_\perp^{\alpha\beta\gamma}(x,x)=2iM_N\left[
N(x,x)g_\perp^{\alpha\beta}\epsilon^{\gamma pnS_\perp}
-N(x,0)(g_\perp^{\beta\gamma}\epsilon^{\alpha pnS_\perp}
+g_\perp^{\gamma\alpha}\epsilon^{\beta pnS_\perp})\right].  
\label{3gluonxx}
\eeq
We remind the derivatives $d/dp_c^\gamma$ and $d/d(x'p'^\gamma)$ in (\ref{master2}) also
hit $\delta\left(\tilde{s}+\tilde{t}+\tilde{u}\right)$.  
The relation (\ref{master2}) shows that the partonic hard cross section for
$O(x,x)$ and $O(x,0)$, in general, differ from each other, and likewise for $N(x,x)$ and $N(x,0)$.

\subsection{Contribution from $O(x,x)$ and $N(x,x)$}

We first consider the contribution in (\ref{master2})
occuring from $O(x,x)$ and $N(x,x)$ in (\ref{3gluonxx}).  
Using the functions (\ref{Hfunction}), 
we write the corresponding hard part as
\beq
&&H^{(F,j)}_{\alpha\beta}(xp,x'p',p_c)g_{\perp}^{\alpha\beta}\epsilon^{\gamma pnS_\perp}
\equiv K^{(F,j)}(\tilde{s},\tilde{t},\tilde{u},m_c^2)\epsilon^{\gamma pnS_\perp},\nonumber\\
&&H^{(I,j)}_{\alpha\beta}(xp,x'p',p_c)g_{\perp}^{\alpha\beta}\epsilon^{\gamma pnS_\perp}
\equiv K^{(I,j)}(\tilde{s},\tilde{t},\tilde{u},m_c^2)\epsilon^{\gamma pnS_\perp},
\label{Kfunction}
\eeq
for $j=d,f$, where we have used the fact that
the scalar functions $K^{(F,j)}$ and $K^{(I,j)}$ ($j=d,f$)
become the functions of $\tilde{s}$, $\tilde{t}$, $\tilde{u}$ and $m_c^2$.  
For the scalar functions $K^{(F,j)}$, $K^{(I,j)}$ and $\delta\left(\tilde{s}+\tilde{t}+\tilde{u}\right)$, 
one can perform the derivative with respect to
$p_c^\gamma$ and $x'p'^\gamma$ in (\ref{master2}) through that with respect to 
$\tilde{u}$ as
\beq
&&{d\over dp_c^\gamma}K^{(F,j)}(\tilde{s},\tilde{t},\tilde{u},m_c^2)
\delta\left(\tilde{s}+\tilde{t}+\tilde{u}\right)
=-2p_{c\gamma}\left(\tilde{s}\over\tilde{t}\right){\partial \over \partial\tilde{u}}
K^{(F,j)}(\tilde{s},\tilde{t},\tilde{u},m_c^2)\delta\left(\tilde{s}+\tilde{t}+\tilde{u}\right),
\label{scalarF}\\
&&{d\over d(x'p'^\gamma)}K^{(I,j)}(\tilde{s},\tilde{t},\tilde{u},m_c^2)
\delta\left(\tilde{s}+\tilde{t}+\tilde{u}\right)
=-2p_{c\gamma}{\partial \over \partial\tilde{u}}
K^{(I,j)}(\tilde{s},\tilde{t},\tilde{u},m_c^2)\delta\left(\tilde{s}+\tilde{t}+\tilde{u}\right),
\label{scalarI}
\eeq
for $j=d,f$, where we have used the fact that $p_c^+$ and $p'^+$ are the dependent variables 
(as noted after (\ref{masterF}) and (\ref{masterI})), and have set ${p'}_\perp^\gamma\to 0$ after taking the 
derivative.\footnote{For the FSI, one can set ${p'}_\perp^\gamma=0$ from the beginning.}
Using (\ref{scalarF}), one obtains for the FSI contribution with $O(x,x)$ in (\ref{master2}) as
\beq
&&
\int {dx\over x^2}  O(x,x)\
{d \over d p_c^\gamma} H_{\alpha\beta}^{(F,d)}
(\tilde{s},\tilde{t},\tilde{u},m_c^2)
\delta\left(\tilde{s}+\tilde{t}+\tilde{u}\right)
g_\perp^{\alpha\beta}\epsilon^{\gamma pnS_\perp}\nonumber\\
&&\quad= -2\epsilon^{p_c pnS_\perp}
\int {dx\over x^2}  O(x,x)\left({\tilde{s}\over\tilde{t}}\right)
{\partial \over \partial \tilde{u}} K^{(F,d)}
(\tilde{s},\tilde{t},\tilde{u},m_c^2)\delta\left(\tilde{s}+\tilde{t}+\tilde{u}\right)\nonumber\\
&&\quad
=-2\epsilon^{p_c pnS_\perp}\int{dx\over x^2} \left(\tilde{s}\over\tilde{t}\right)\left[
\left( {\partial K^{(F,d)}\over \partial \tilde{u}}
+ {x\over \tilde{u}}{ \partial K^{(F,d)} \over \partial x}
-{2K^{(F,d)}\over \tilde{u}}\right) O(x,x)
+{K^{(F,d)}\over \tilde{u}} x{dO(x,x)\over dx}\right]\nonumber\\
&&\qquad\qquad\qquad\qquad\qquad\times\delta\left(\tilde{s}+\tilde{t}+\tilde{u}\right)\nonumber\\
&&\quad =-2\epsilon^{p_c pnS_\perp}\int{dx\over x^2} \left(\tilde{s}\over\tilde{t}\right)\left[
\left( {\partial K^{(F,d)}\over \partial \tilde{u}}
+{\tilde{t}\over \tilde{u}}{\partial K^{(F,d)}\over \partial \tilde{t}}
+{\tilde{s}\over \tilde{u}}{\partial K^{(F,d)}\over \partial \tilde{s}}
-{2K^{(F,d)}\over \tilde{u}}\right) O(x,x)\right.\nonumber\\
&&\qquad\qquad\qquad\qquad\qquad\qquad\qquad\left.+{K^{(F,d)}\over \tilde{u}} x{dO(x,x)\over dx}\right]
\delta\left(\tilde{s}+\tilde{t}+\tilde{u}\right)\nonumber\\
&&\quad
=-2\epsilon^{p_c pnS_\perp}\int{dx\over x^2} \left(\tilde{s}\over\tilde{t}\tilde{u}\right)\left[ 
- m_c^2 {\partial K^{(F,d)}\over \partial m_c^2} O(x,x)
+K^{(F,d)}\left( x{dO(x,x)\over dx}-2O(x,x)\right) 
\right]\nonumber\\
&&\qquad\qquad\qquad\qquad\qquad\qquad\times
\delta\left(\tilde{s}+\tilde{t}+\tilde{u}\right).  
\label{scalarFSI}
\eeq
In the first equality of (\ref{scalarFSI}), we transformed the derivative hitting 
$\delta\left(\tilde{s}+\tilde{t}+\tilde{u}\right)$ into the derivative with respect to $x$ and performed
the partial integration.  In the last equality
we have used the relation
\beq
\left(
\tilde{s}{\partial \over \partial \tilde{s}} +
\tilde{t}{\partial \over \partial \tilde{t}} +
\tilde{u}{\partial \over \partial \tilde{u}} +
m_c^2{\partial \over \partial m_c^2} \right)K^{(B,j)}(\tilde{s},\tilde{t},\tilde{u},m_c^2)=0,
\eeq
for $B=F, I$ and $j=d, f$, resulting from the  
scale-invariance property for the dimensionless function
$K^{(B,j)}(\tilde{s},\tilde{t},\tilde{u},m_c^2)=
K^{(B,j)}(\lambda\tilde{s},\lambda\tilde{t},\lambda\tilde{u},\lambda m_c^2)$.  
Similarly to (\ref{scalarFSI}), by using (\ref{scalarI}), 
one obtains for the ISI contribution with $O(x,x)$ in (\ref{master2}) as
\beq
&&
\hspace{-0.7cm}-\int {dx\over x^2}  O(x,x)\
{d \over d (x'p'^\gamma)} H_{\alpha\beta}^{(I,d)}
(\tilde{s},\tilde{t},\tilde{u},m_c^2)
\delta\left(\tilde{s}+\tilde{t}+\tilde{u}\right)
g_\perp^{\alpha\beta}\epsilon^{\gamma pnS_\perp}\nonumber\\
&&\hspace{-0.5cm}=2\epsilon^{p_c pnS_\perp}
\int {dx\over x^2}  O(x,x) 
{\partial \over \partial \tilde{u}} K^{(I,d)}
(\tilde{s},\tilde{t},\tilde{u},m_c^2)\delta\left(\tilde{s}+\tilde{t}+\tilde{u}\right)\nonumber\\
&&\hspace{-0.5cm}
=2\epsilon^{p_c pnS_\perp}\int{dx\over x^2}
{1\over \tilde{u}}\left[ 
- m_c^2 {\partial K^{(I,d)}\over \partial m_c^2} O(x,x)
+ K^{(I,d)} \left( x{dO(x,x)\over dx} -2O(x,x)\right)
\right]
\delta\left(\tilde{s}+\tilde{t}+\tilde{u}\right).  \nonumber\\
\label{scalarISI}
\eeq
By the replacement 
$K^{(B,d)}(\tilde{s},\tilde{t},\tilde{u},m_c^2)
\to K^{(B,f)}(\tilde{s},\tilde{t},\tilde{u},m_c^2)$ ($B=F,I$) and $O(x,x) \to N(x,x)$ in
(\ref{scalarFSI}) and (\ref{scalarISI}), one obtains the formula
for the $N(x,x)$ contributions.  
From (\ref{scalarFSI}) and (\ref{scalarISI}), one can make the following important observations
for the $O(x,x)$ and $N(x,x)$ contributions:  
\begin{enumerate}
\item[(1)] The partonic hard cross sections for
$x{d\over dx}O(x,x)$ and $O(x,x)$ are connected by the simple relation.
In particular, in the $m_c\to 0$ limit, they contribute in the form
$x{d\over dx}O(x,x) - 2O(x,x)$.  The same relation holds for 
$x{d\over dx}N(x,x)$ and $N(x,x)$.  

\item[(2)] The partonic hard cross sections $K^{(B,j)}$ ($B=F,I$, $j=d,f$)
defined in (\ref{Kfunction})
are obtained from ${\cal H}_{\mu\nu}^{F,abc}$ and ${\cal H}_{\mu\nu}^{I,abc}$ 
by the same Lorentz contraction as the unpolarized cross section
in (\ref{unpolH}).  Therefore the contribution to $K^{(B,j)}$ from each diagram
differs from those for $\hat{\sigma}^U_{gg\to c}$ only in the color factors.  
\end{enumerate}
These features are the extension of
those obtained in \cite{KT072} for the SGP contribution
of the quark-gluon correlation function with massless partons 
to the case of the three-gluon correlation functions with massive partons in the final state.  

\subsection{Contribution from $O(x,0)$ and $N(x,0)$}

Next we consider the contribution from $O(x,0)$ and $N(x,0)$ in (\ref{master2}), which arise 
from the second terms in (\ref{3gluonxx}).  
For this purpose, we introduce the two fixed vectors
$X^\mu =(0,1,0,0)$ and $Y^\mu =(0,0,1,0)$, and write
\beq
g_{\perp}^{\beta\gamma}= -X^\beta X^\gamma -Y^\beta Y^\gamma.  
\eeq
Then the derivative ${d/ dp_c^\gamma}$ hitting the FSI hard part in (\ref{master2})
for $O(x,0)$ and $N(x,0)$ can be written as
\beq
&&
{d\over dp_c^\gamma} H_{\alpha\beta}^{(F,j)}(xp,x'p',p_c)\delta\left(\tilde{s}+\tilde{t}+\tilde{u}\right)
\left( g_\perp^{\beta\gamma}\epsilon^{\alpha pnS_\perp} + g_\perp^{\alpha\gamma}\epsilon^{\beta pnS_\perp}\right)
\nonumber\\
&&\qquad
=-X^\mu{d\over dp_c^\mu} H_{\alpha\beta}^{(F,j)}(xp,x'p',p_c)\delta\left(\tilde{s}+\tilde{t}+\tilde{u}\right)
\left( X^{\beta}\epsilon^{\alpha pnS_\perp} + X^{\alpha}\epsilon^{\beta pnS_\perp}\right)\nonumber\\
&&\qquad\quad -Y^\mu{d\over dp_c^\mu} H_{\alpha\beta}^{(F,j)}(xp,x'p',p_c)
\delta\left(\tilde{s}+\tilde{t}+\tilde{u}\right)
\left( Y^{\beta}\epsilon^{\alpha pnS_\perp} + Y^{\alpha}\epsilon^{\beta pnS_\perp}\right). 
\label{deriv}
\eeq
To perform the derivatives in this equation, we introduce the scalar
functions $J_{1,2}^{(F,j)}(\tilde{s},\tilde{t},\tilde{u})$ ($j=d,\,f$) by the decomposition:
\beq
&&H^{(F,j)}_{\alpha\beta}(xp,x'p',p_c) 
\left(X^\beta \epsilon^{\alpha pnS_\perp} + X^\alpha \epsilon^{\beta pnS_\perp}\right)\nonumber\\
&&\qquad\qquad\equiv 
J_1^{(F,j)}(\tilde{s},\tilde{t},\tilde{u},m_c^2)\left(p_c\cdot X\right) \epsilon^{p_c pnS_\perp}
+J_2^{(F,j)}(\tilde{s},\tilde{t},\tilde{u},m_c^2)  \epsilon^{X pnS_\perp},
\label{JFSI}
\eeq
where we used the kinematic condition $p_\perp^\gamma={p'}_\perp^\gamma=0$.   
One also obtains  
the similar decomposition for $H^{(F,j)}_{\alpha\beta}(xp,x'p',p_c) 
\left(Y^\beta \epsilon^{\alpha pnS_\perp} + Y^\alpha \epsilon^{\beta pnS_\perp}\right)$
by the replacement $X\to Y$ in (\ref{JFSI}) 
with the same functions $J_{1,2}^{(F,j)}$.  
Then the derivative in (\ref{deriv}) can be performed 
with the help of (\ref{scalarF}) as
\beq
&&-X^\mu{d\over dp_c^\mu}\left[ J_1^{(F,j)}(\tilde{s},\tilde{t},\tilde{u},m_c^2)
(p_c\cdot X)\epsilon^{p_cpnS_\perp}
\delta\left(\tilde{s}+\tilde{t}+\tilde{u}\right)
\right] +(X\to Y)\nonumber\\
&&\qquad\qquad=3\epsilon^{p_cpnS_\perp}J_1^{(F,j)}(\tilde{s},\tilde{t},\tilde{u},m_c^2)
\delta\left(\tilde{s}+\tilde{t}+\tilde{u}\right)\nonumber\\
&&\qquad\qquad\quad+2\vec{p}_{c\perp}^{\ 2}\epsilon^{p_cpnS_\perp}\left({\tilde{s}\over \tilde{t}}\right)
{\partial \over \partial\tilde{u}}
\left(J_1^{(F,j)}(\tilde{s},\tilde{t},\tilde{u},m_c^2)
\delta\left(\tilde{s}+\tilde{t}+\tilde{u}\right)\right),\\
&&-X^\mu{d\over dp_c^\mu}\left[ J_2^{(F,j)}
(\tilde{s},\tilde{t},\tilde{u},m_c^2)\epsilon^{XpnS_\perp}
\delta\left(\tilde{s}+\tilde{t}+\tilde{u}\right)
\right] +(X\to Y)\nonumber\\
&&\qquad\qquad=-2\epsilon^{p_cpnS_\perp}{\tilde{s}\over \tilde{t}}
{\partial \over \partial\tilde{u}}
\left(J_2^{(F,j)}(\tilde{s},\tilde{t},\tilde{u},m_c^2)
\delta\left(\tilde{s}+\tilde{t}+\tilde{u}\right)\right).  
\eeq
Using these results and the relation
$\vec{p}_{c\perp}^{\ 2}={\tilde{t}\tilde{u}\over \tilde{s}}-m_c^2$ in (\ref{deriv}), and
following the same procedure leading to (\ref{scalarFSI}), 
one 
obtains the FSI contribution with $O(x,0)$ and $N(x,0)$ in (\ref{master2}) 
in terms of 
$J_{1,2}^{(F,j)}$ in (\ref{JFSI}) as 
\beq
&&\int{dx\over x^2}
{d\over dp_c^\gamma} H_{\alpha\beta}^{(F,d)}(xp,x'p',p_c) \delta\left(\tilde{s}+\tilde{t}+\tilde{u}\right)
\left( g_\perp^{\beta\gamma}\epsilon^{\alpha pnS_\perp} + 
g_\perp^{\alpha\gamma}\epsilon^{\beta pnS_\perp}\right)O(x,0)\nonumber\\
&&\qquad\qquad+\left( H_{\alpha\beta}^{(F,d)}\to H_{\alpha\beta}^{(F,f)},\ O(x,0) \to -N(x,0)\right)
\nonumber\\
&&\quad=\epsilon^{p_c pnS_\perp}\int{dx\over x^2}\left[
3J_1^{(F,d)}
O(x,0) 
+{2\tilde{s}\over \tilde{t}\tilde{u}}
\left( {\tilde{t}\tilde{u}\over \tilde{s}} -m_c^2\right)
\right.\nonumber\\
&&\qquad\quad\left. \times\left\{
\left( -J_1^{(F,d)} -m_c^2{\partial J_1^{(F,d)}\over \partial m_c^2}\right)O(x,0)
+J_1^{(F,d)}\left( x{d O(x,0)\over dx} -2O(x,0)\right)
\right\}
\right.\nonumber\\
&&
\left.
\qquad
- {2\tilde{s}\over \tilde{t}\tilde{u}}
\left\{
-m_c^2{\partial J_2^{(F,d)}\over \partial m_c^2}O(x,0)
+J_2^{(F,d)}\left( x{d O(x,0)\over dx} -2O(x,0)\right)
\right\}
\right]\delta\left(\tilde{s}+\tilde{t}+\tilde{u}\right)\nonumber\\
&&\qquad +\left( J_{1,2}^{(F,d)} \to J_{1,2}^{(F,f)},\  O(x,0)\to -N(x,0)\right). 
\label{FSIx0}
\eeq

The ISI contribution in (\ref{master2}) 
with $O(x,0)$ and $N(x,0)$ can also be obtained following the same procedure
as above.  Similarly to (\ref{JFSI}),
one can decompose 
the ISI hard part as 
\beq
&&\hspace{-1cm}H^{(I,j)}_{\alpha\beta}(xp,x'p',p_c) 
\left(X^\beta \epsilon^{\alpha pnS_\perp} + X^\alpha \epsilon^{\beta pnS_\perp}\right)\nonumber\\
&&\hspace{-0.3cm}\equiv 
J_1^{(I,j)}(\tilde{s},\tilde{t},\tilde{u},m_c^2)\left(p_c\cdot X\right) \epsilon^{p_c pnS_\perp}
+J_2^{(I,j)}(\tilde{s},\tilde{t},\tilde{u},m_c^2)  \epsilon^{X pnS_\perp}\nonumber\\
&&+J_3^{(I,j)}(\tilde{s},\tilde{t},\tilde{u},m_c^2)\left(x'p'\cdot X\right) \epsilon^{p_c pnS_\perp}
+J_4^{(I,j)}(\tilde{s},\tilde{t},\tilde{u},m_c^2)\left(p_c\cdot X\right) x'\epsilon^{p' pnS_\perp}
\label{JISI}
\eeq
for $j=d,f$, where we ignored the terms which vanish in the limit ${p'}_\perp^\gamma\to 0$
after taking the derivative $d/d(x'p'^\gamma)$.  
Similar relation can be written down for\\
$H^{(I,j)}_{\alpha\beta}(xp,x'p',p_c) 
\left(Y^\beta \epsilon^{\alpha pnS_\perp} + Y^\alpha \epsilon^{\beta pnS_\perp}\right)$,
using the same functions $J_{1,2,3,4}^{(I,j)}$.  
Compared with (\ref{JFSI}),
note the existence of the $J_{3,4}^{(I,j)}$ terms in (\ref{JISI}), since one has to take
the derivative with respect to ${p'}_\perp^\gamma$ before taking the ${p'}_\perp^\gamma \to 0$ limit.  
With these $J_{1,2,3,4}^{(I,j)}$, 
one eventually obtains the ISI contribution in (\ref{master2}) with
$O(x,0)$ and $N(x,0)$ as
\beq
&&-\int{dx\over x^2}
{d\over d(x'p'^\gamma)} H_{\alpha\beta}^{(I,d)}(xp,x'p',p_c) 
\delta\left(\tilde{s}+\tilde{t}+\tilde{u}\right)
\left( g_\perp^{\beta\gamma}\epsilon^{\alpha pnS_\perp} + 
g_\perp^{\alpha\gamma}\epsilon^{\beta pnS_\perp}\right)O(x,0)\nonumber\\
&&\qquad\qquad+\left( H_{\alpha\beta}^{(I,d)}\to H_{\alpha\beta}^{(I,f)},\ O(x,0) \to -N(x,0)\right)
\nonumber\\
&&\quad=\epsilon^{p_cpnS_\perp}\int{dx\over x^2}
\left[
{2\over \tilde{u}}
\left( {\tilde{t}\tilde{u}\over \tilde{s}} -m_c^2\right)
\left\{
\left( J_1^{(I,d)} +m_c^2{\partial J_1^{(I,d)}\over \partial m_c^2}\right)O(x,0)
\right.\right.\nonumber\\
&&\left.\left. \qquad\qquad\qquad
\qquad\qquad\qquad\qquad -J_1^{(I,d)}\left( x{d O(x,0)\over dx} -2O(x,0)\right)\right\}
\right.\nonumber\\
&&\left.\qquad\qquad\qquad
+ {2\over \tilde{u}}
\left\{
-m_c^2{\partial J_2^{(I,d)}\over \partial m_c^2}O(x,0)
+J_2^{(I,d)}\left( x{d O(x,0)\over dx} -2O(x,0)\right)
\right\}
\right.\nonumber\\
&&\left.\qquad\qquad\qquad\qquad\qquad
-(2J_3^{(I,d)}+J_4^{(I,d)})O(x,0)
\right]
\delta\left(\tilde{s}+\tilde{t}+\tilde{u}\right)\nonumber\\
&&\qquad +\left( J_{1,2,3,4}^{(I,d)} \to J_{1,2,3,4}^{(I,f)},\  O(x,0)\to -N(x,0)\right).  
\label{ISIx0}
\eeq

\subsection{Twist-3 cross section from the $gg\to c\bar{c}$ scattering}

Using the results 
(\ref{scalarFSI}), (\ref{scalarISI}), (\ref{FSIx0}) and (\ref{ISIx0})
in (\ref{master2}), one 
obtains the final result for the total twist-3 single-spin-dependent 
cross section for $p^\uparrow p\to DX$ induced by the
three-gluon correlation functions as 
\beq 
&&\hspace{-0.7cm}P_h^0\frac{d\Delta\sigma}{d^3P_h}=\frac{2\pi M_N\alpha_s^2}{S}\epsilon^{P_hpnS_\perp}
\int\frac{dx'}{x'}G(x')\int\frac{dz}{z^3}D_c(z)\int\frac{dx}{x^2}
\nonumber\\
&&\qquad\times\left[\left(2\tilde{s}\over\tilde{t}\tilde{u}\right)\left\{
m_c^2 {\partial K^{(F,d)}\over \partial m_c^2} O(x,x)
-K^{(F,d)}\left( x{dO(x,x)\over dx}-2O(x,x)\right) 
\right\}\right.\nonumber\\
&&\left.\qquad\qquad+{2\over \tilde{u}}\left\{
- m_c^2 {\partial K^{(I,d)}\over \partial m_c^2} O(x,x)
+ K^{(I,d)} \left( x{dO(x,x)\over dx} -2O(x,x)\right)
\right\}
\right.\nonumber\\
&&\left.\qquad\qquad+\left( O(x,x)\to N(x,x),\ \ K^{(F,d)} \to K^{(F,f)},\ \ K^{(I,d)} \to K^{(I,f)}
\right)\right.\nonumber\\
&&\left.
\qquad+3J_1^{(F,d)}
O(x,0) 
+{2\tilde{s}\over \tilde{t}\tilde{u}}
\left( {\tilde{t}\tilde{u}\over \tilde{s}} -m_c^2\right)
\right.\nonumber\\
&&\qquad\quad\left. \times\left\{
\left( -J_1^{(F,d)} -m_c^2{\partial J_1^{(F,d)}\over \partial m_c^2}\right)O(x,0)
+J_1^{(F,d)}\left( x{d O(x,0)\over dx} -2O(x,0)\right)
\right\}
\right.\nonumber\\
&&
\left.
\qquad\qquad
- {2\tilde{s}\over \tilde{t}\tilde{u}}
\left\{
-m_c^2{\partial J_2^{(F,d)}\over \partial m_c^2}O(x,0)
+J_2^{(F,d)}\left( x{d O(x,0)\over dx} -2O(x,0)\right)
\right\}\right.\nonumber\\
&&\left.\qquad+\left( O(x,0)\to -N(x,0),\ \ J_{1,2}^{(F,d)} \to J_{1,2}^{(F,f)}\right)\right.\nonumber\\
&&\left.+{2\over \tilde{u}}
\left( {\tilde{t}\tilde{u}\over \tilde{s}} -m_c^2\right)
\left\{
\left( J_1^{(I,d)} +m_c^2{\partial J_1^{(I,d)}\over \partial m_c^2}\right)O(x,0)
-J_1^{(I,d)}\left( x{d O(x,0)\over dx} -2O(x,0)\right)\right\}
\right.\nonumber\\
&&\left.
+ {2\over \tilde{u}}
\left\{
-m_c^2{\partial J_2^{(I,d)}\over \partial m_c^2}O(x,0)
+J_2^{(I,d)}\left( x{d O(x,0)\over dx} -2O(x,0)\right)
\right\}
-(2J_3^{(I,d)}+J_4^{(I,d)})O(x,0)\right.\nonumber\\
&&\left.\qquad +\left( O(x,0)\to -N(x,0),\ \ J_{1,2,3,4}^{(I,d)} \to J_{1,2,3,4}^{(I,f)}\right) 
\right]
\delta\left(\tilde{s}+\tilde{t}+\tilde{u}\right).
\label{masterfinal}
\eeq
where $K^{(B,j)}(\tilde{s},\tilde{t},\tilde{u},m_c^2)$ ($B=F, I$, $j=d, f$),
$J^{(F,j)}_{1,2}(\tilde{s},\tilde{t},\tilde{u},m_c^2)$ ($j=d, f$) and
$J^{(I,j)}_{1,2,3,4}(\tilde{s},\tilde{t},\tilde{u},m_c^2)$ ($j=d, f$) are the functions
defined, respectively, in (\ref{Kfunction}),
(\ref{JFSI}) and (\ref{JISI})
and can be calculated from the twist-2 diagrams in Fig. 2.  
By the direct calculation of
these functions, we found that $J_1^{(F,j)}$ and $J_{1,3,4}^{(I,j)}$ ($j=d,f$)
are of $O(m_c^2)$ and thus vanish in the $m_c\to 0$ limit, and that
$K^{(B,j)}=J_2^{(B,j)}$ ($B=F,I$, $j=d,f$) at $m_c= 0$, which is consistent with 
the result in (\ref{twist3final}). 
For the twist-3 cross section for the $\bar{D}$-meson production,
the sign of the contribution from $O(x,x)$ and $O(x,0)$ should be reversed in (\ref{masterfinal}). 
The result calculated from (\ref{masterfinal}), of course, agrees with (\ref{twist3final})
which were obtained by the direct calculation of Fig. 4.  

The origin of the above master formula (\ref{masterfinal}) 
is the relations in (\ref{masterF}) and (\ref{masterI}).  Although they were derived in the LO QCD,  
the derivation was based on the quite general structure of the diagrams for the SGP contribution
at $x_1=x_2$ shown in Fig. 3, i.e., the coherent-gluon line is attached to the external
parton line of the twist-2 diagrams in Fig. 1.
As long as this structure is kept after including higher-order corrections, 
the relations (\ref{masterF}) and (\ref{masterI}) hold.  
Therefore we expect that the formula (\ref{masterfinal}) will become a powerful tool
to include high-order corrections to the twist-3 SSA\,\cite{Koike:2011ns}.

\section{Numerical calculation of the asymmetry}

As is shown in (\ref{twist3final}), four nonperturbative functions 
$O(x,x)$, $N(x,x)$, $O(x,0)$ and $N(x,0)$  
participate in the twist-3 cross section for $A_N^D$.  
Unlike twist-2 parton distributions, twist-3 multiparton correlation functions
do not have probability interpretation and thus
cannot be constrained by a certain positivity bound.  
\footnote{For a comprehensive review on the positivity bounds of 
parton distributions, see \cite{inequality}.}
They have to be determined by comparing
the calculated SSAs with experimental data,
or by some nonperturbative techniques in QCD.  
At present there is no information on these functions.  
Preliminary data on $A_N^D$ by the PHENIX collaboration\,\cite{Liu} suggests $|A_N^D|\leq 5$ \%
in the region $|x_F|<0.1$ at $\sqrt{S}=200$ GeV.  
Since the unpolarized cross section for the pion production at RHIC
has been well described by the next-to-leading order (NLO) calculation in the collinear 
factorization\,\cite{JagerSchaferStratmannVogelsang2003,Star2004,Phenix2005},
comparison of the asymmetry calculated by our twist-3 cross section 
with the RHIC data will be the first step for determining
the magnitude of the three-gluon correlation
functions in the nucleon.
\footnote{The size of the NLO correction to the twist-3 cross section in (\ref{twist3final})
could be different from that for the twist-2 unpolarized cross section, which may
lead to significantly different $A_N^D$.  
One thus should take the present calculation as only an estimate of the order-of-magnitude.}
Here we present a simple model calculation of the asymmetry at the RHIC energy
taking into account of the preliminary RHIC data.  

\begin{figure}[h]
\begin{center}

\vspace{-0.5cm}
\scalebox{0.6}{\includegraphics{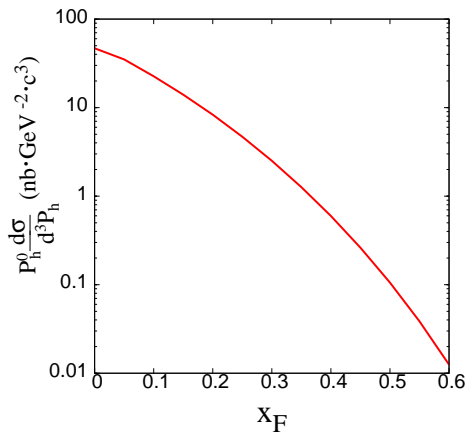}}

\caption{Unpolarized cross section for $pp\to DX$ 
by the gluon fusion process at the RHIC energy
$\sqrt{S}=200$ GeV and $P_T=2$ GeV. }
\end{center}
\end{figure}

To see the relative importance of each term appearing in (\ref{twist3final}),
we assume the same form for the four nonperturbative functions as
\footnote{Minus sign is introduced for $N(x,0)$, since 
$\hat{\sigma}^{N2}$ has an opposite sign compared with $\hat{\sigma}^{O1,O2,N1}$.
}
\beq
O(x,x)=O(x,0)=N(x,x)=-N(x,0).
\label{OONN}
\eeq
As a functional form of these functions, 
we employ the following ansatz:  
\beq
&&{\rm Model\ 1}:\qquad O(x,x)=K_G\,x\,G(x),
\label{case1}\\
&&{\rm Model\ 2}:\qquad O(x,x)=K_G'\,\sqrt{x}\,G(x), 
\label{case2}
\eeq
where $G(x)$ is the twist-2 unpolarized gluon density, and $K_G$ and $K_G'$ are the constants 
which we determine so that the calculated asymmetry is consistent with the RHIC data.
\begin{figure}[h]
\begin{center}

\vspace{-0.5cm}
\scalebox{0.6}{\includegraphics{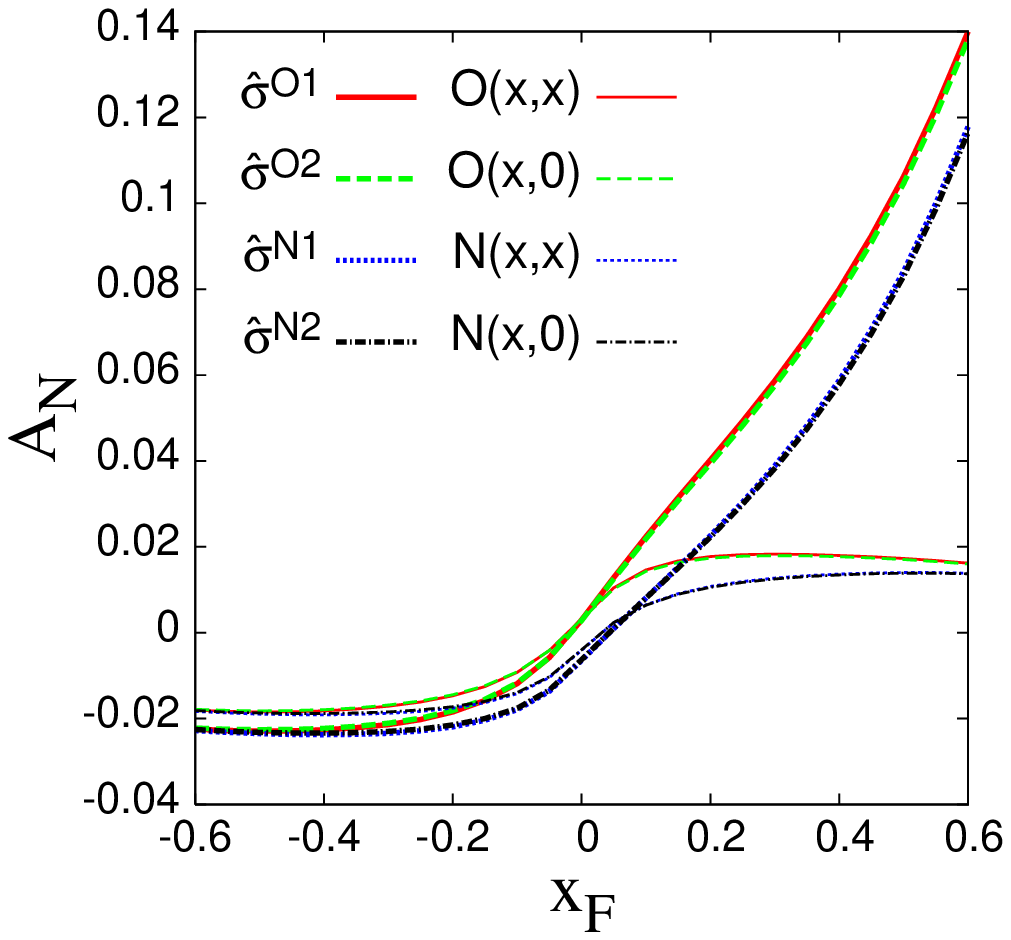}}
\scalebox{0.6}{\includegraphics{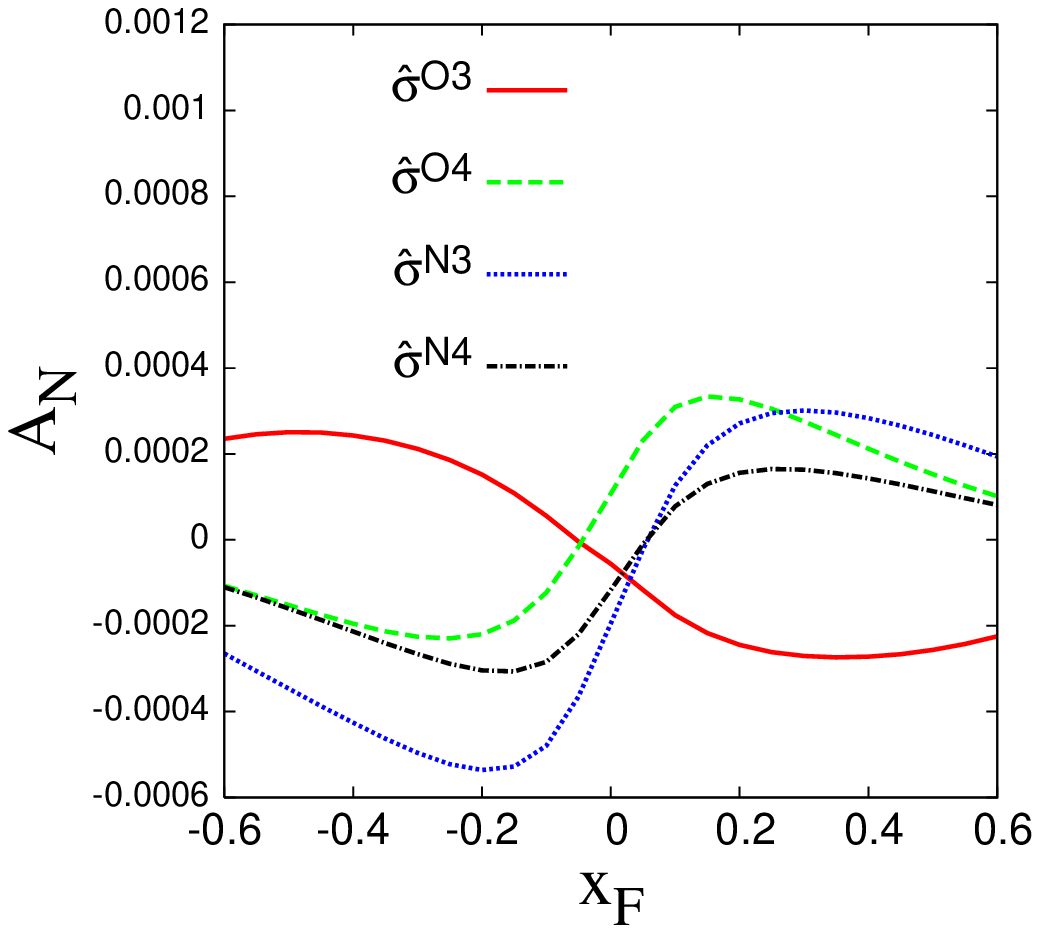}}

\vspace{-0.3cm}
(a)

\vspace{0.3cm}

\scalebox{0.6}{\includegraphics{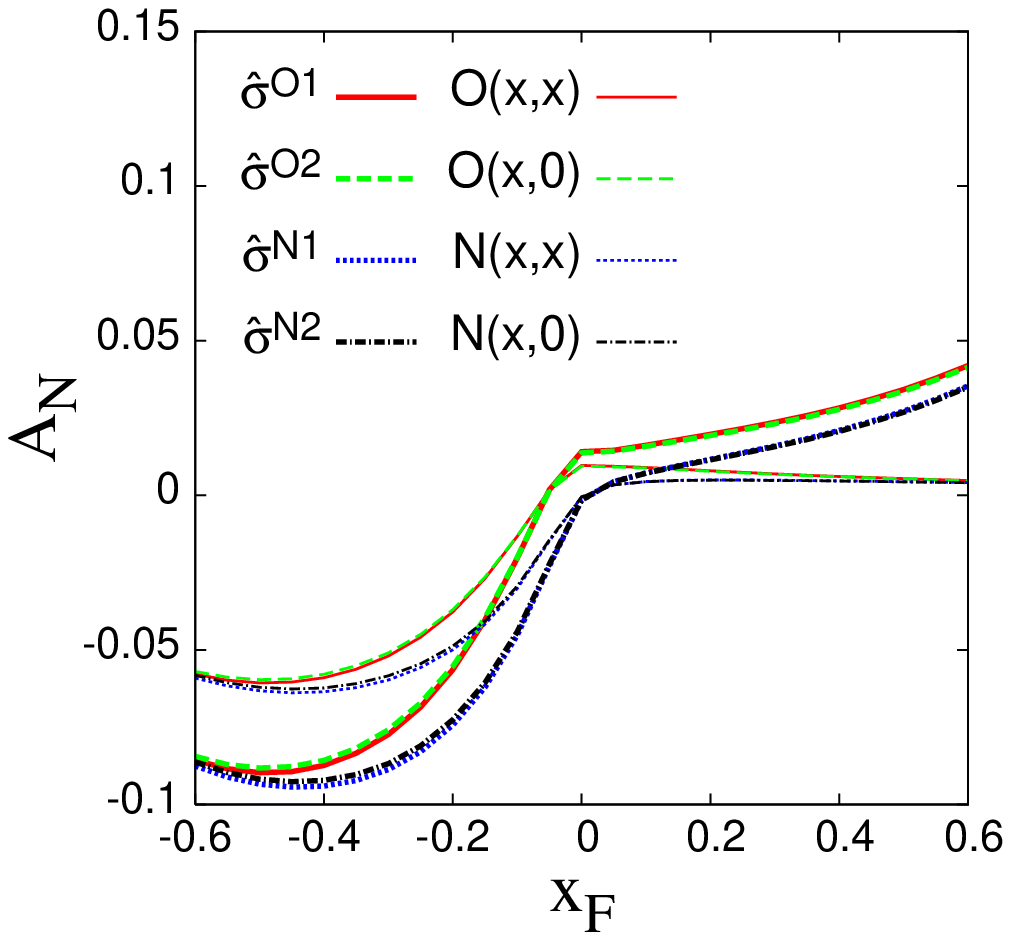}}
\scalebox{0.6}{\includegraphics{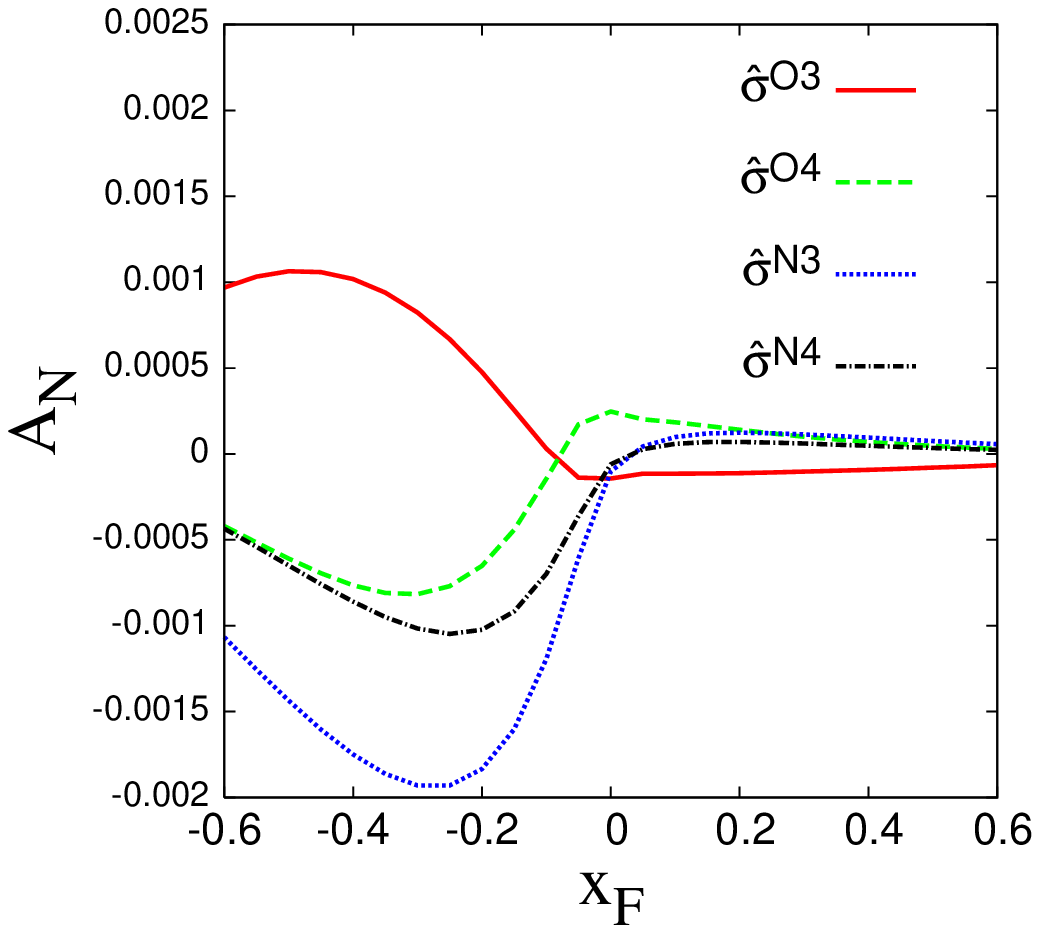}}

(b)

\caption{(a) Contribution to $A_N^D$ from the 8 components proportional to
$\sigma^{O1,O2,N1,N2}$ (left) and $\sigma^{O3,O4,N3,N4}$ (right) in (\ref{twist3final})
obtained by using the model 1 in (\ref{case1}) with $K_G=0.002$.
In the left figure, the non-derivative terms contributing with $\sigma^{O1,O2,N1,N2}$
are also plotted by thin lines labeled by $O(x,x)$, $O(x,0)$, $N(x,x)$ and $N(x,0)$. 
(b) The same as (a) but for the model 2 in (\ref{case2}) with $K_G'=0.0005$. 
}
\end{center}
\end{figure}
Since the three-gluon correlation functions
are completely independent from the gluon density, 
the above parametrization 
is a very crude
approximation and the result below should be taken 
only as an estimate of the order-of-magnitude for the three-gluon correlation functions. 
But they are useful to get the 
shape and magnitude of the three-gluon correlation functions relative to
the gluon density.  
Note that the above two models monitor the sensitivity of $A_N^D$ to the small-$x$ behavior of the
three-gluon correlation functions.  
For the numerical calculation, we use GJR08 distribution\,\cite{GJR08} for $G(x)$ and 
KKKS08 fragmentation function\,\cite{KKKS08} for $D_f(z)$.  
We also assume the same scale dependence for $O(x,x)$ {\it etc} as $G(x)$
for simplicity.  
We calculate $A_N$ for the $D$ and $\bar{D}$ mesons 
at the RHIC energy of $\sqrt{S}=200$ GeV and the transverse momentum of the $D$-meson
$P_T=2$ GeV
with the parameter $m_c=1.3$ GeV by setting the 
scale of all the distribution and fragmentation functions at $\mu=\sqrt{P_T^2+m_c^2}$.

For completeness, we first show in Fig. 5 the unpolarized cross section
for $pp\to DX$ based on the gluon fusion process (\ref{unpol})
at $\sqrt{S}=200$ GeV and $P_T=2$ GeV.  This will constitute the denominator
of $A_N^D$ in our calculation below.

\begin{figure}[h]
\begin{center}

\vspace{-0.5cm}
\scalebox{0.6}{\includegraphics{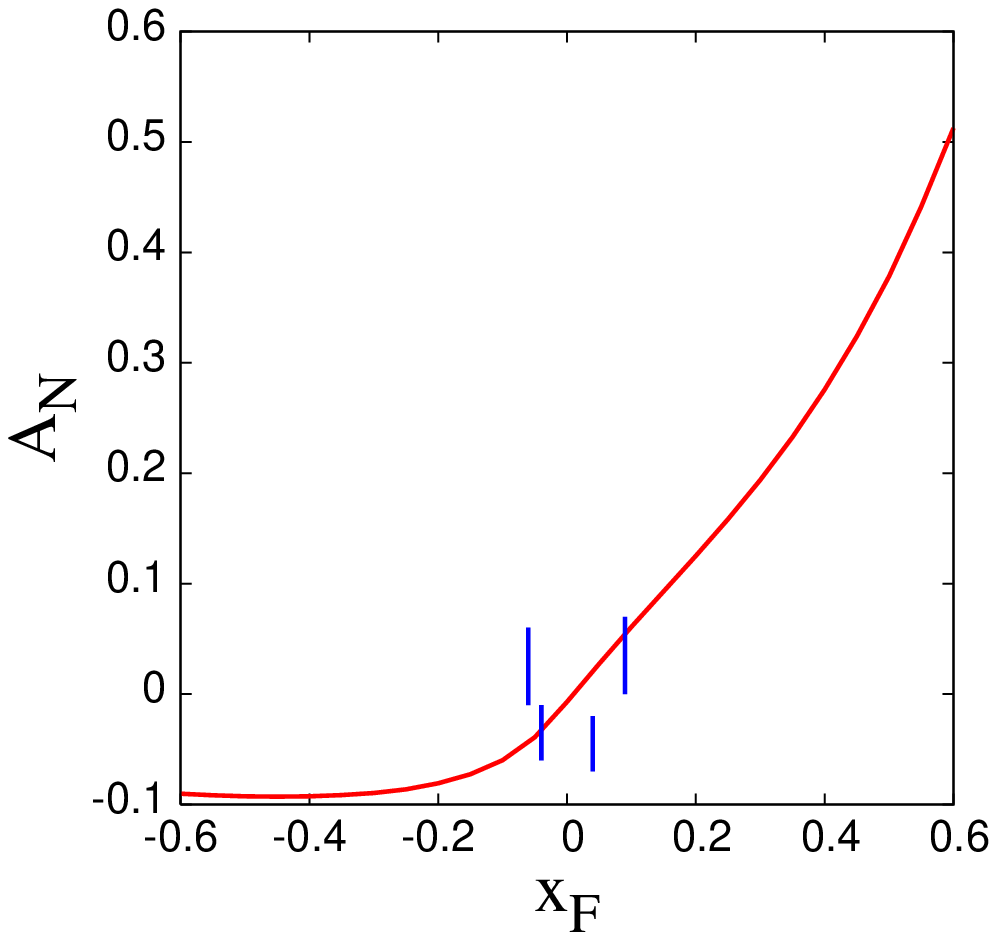}}
\scalebox{0.6}{\includegraphics{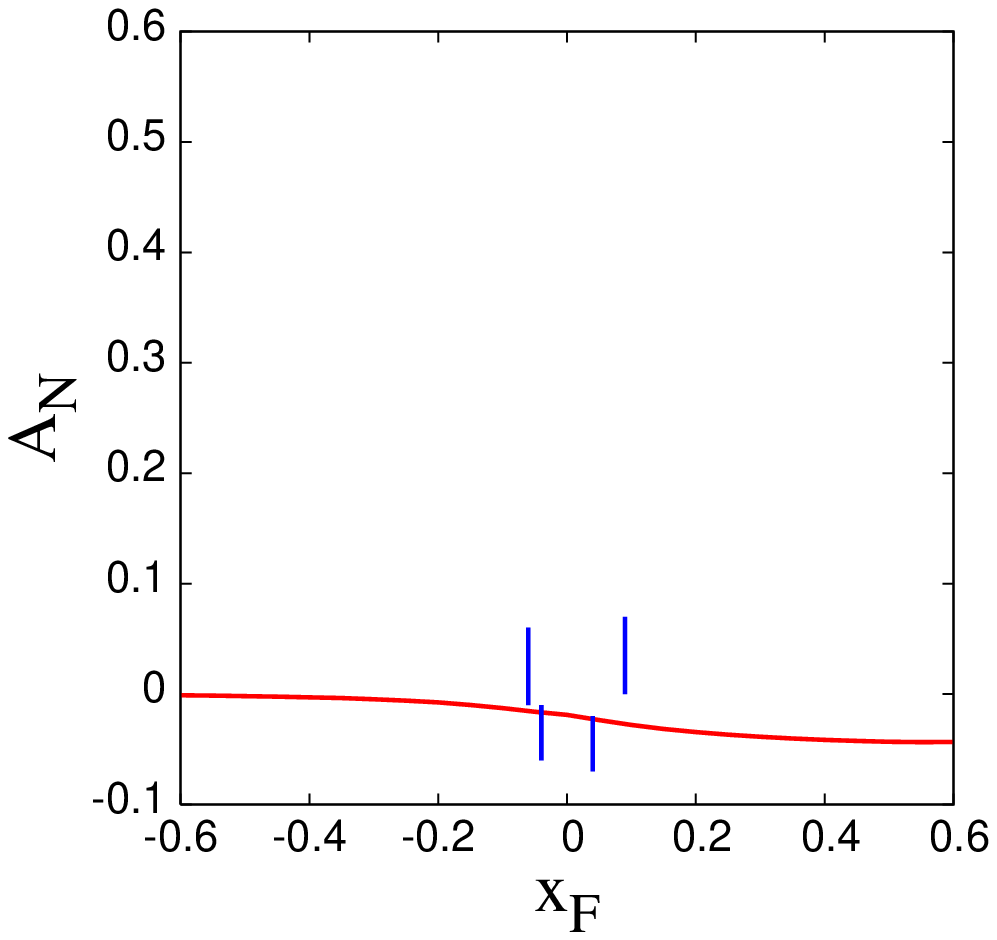}}

\vspace{-0.3cm}
(a)

\vspace{0.3cm}

\scalebox{0.6}{\includegraphics{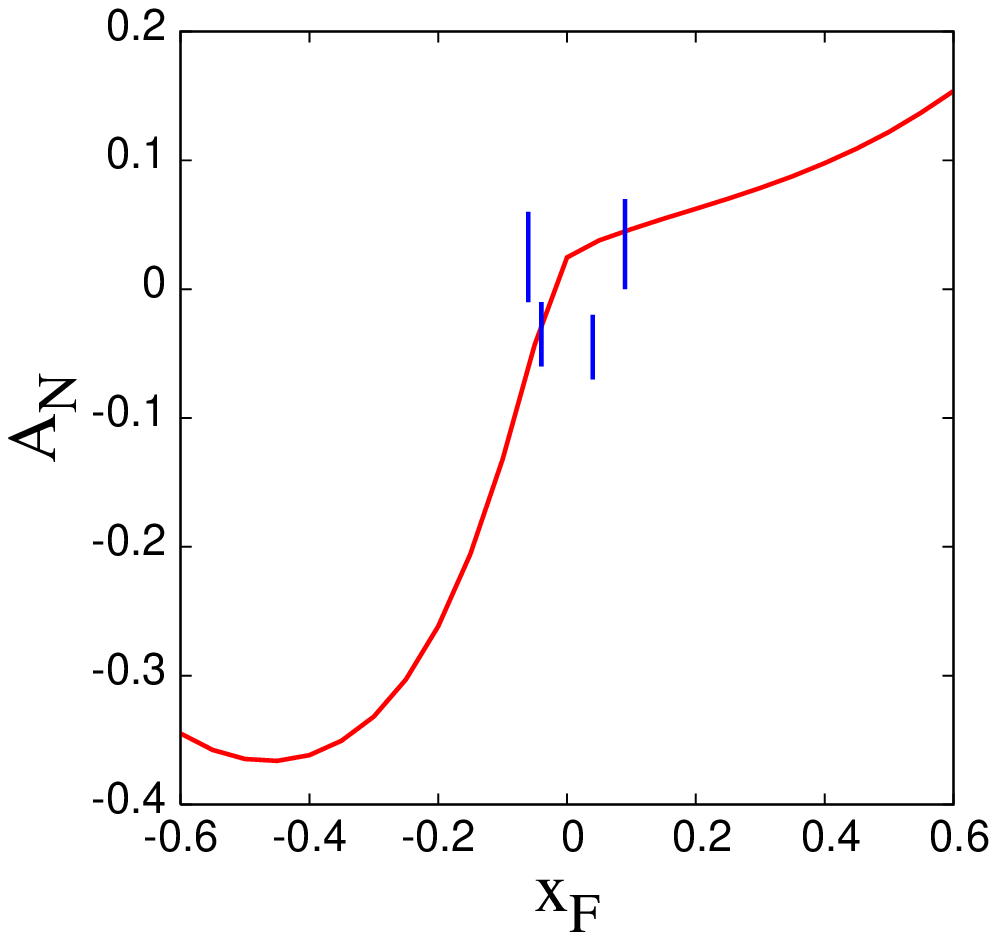}}
\scalebox{0.6}{\includegraphics{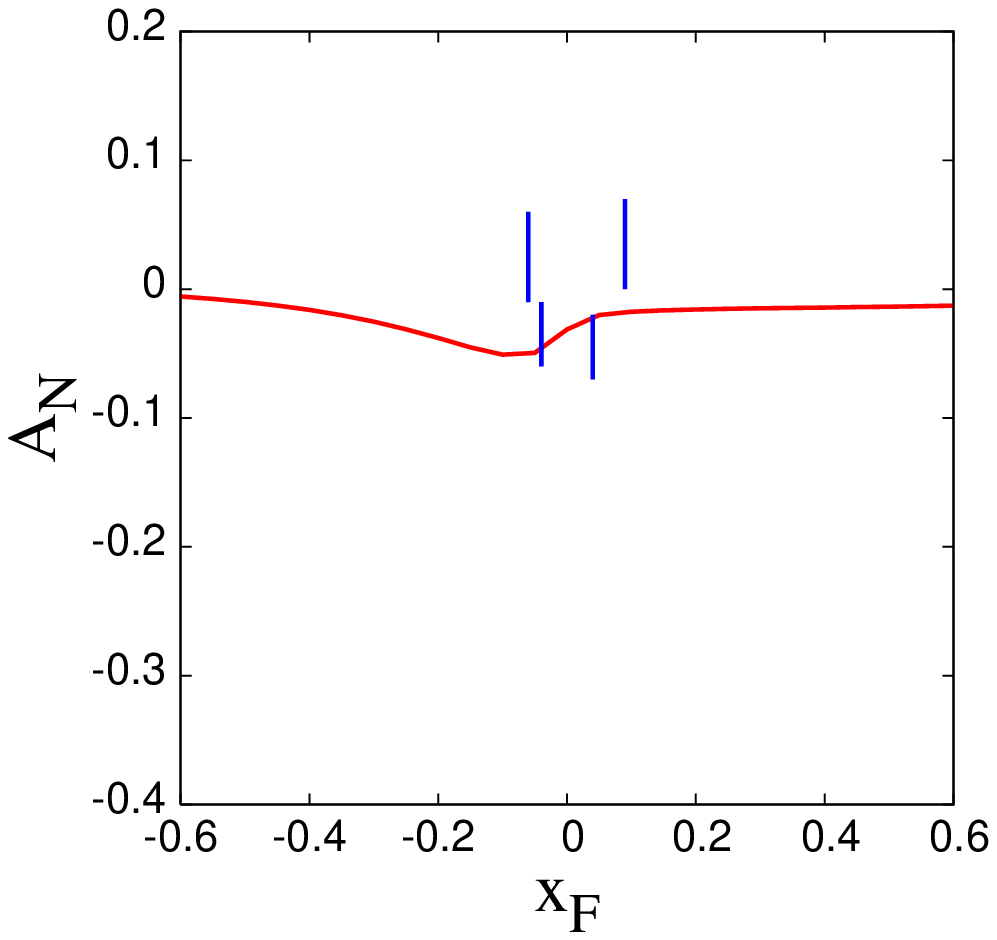}}

(b)

\caption{(a) $A_N^D$ for the $D^0$ (left) and $\bar{D^0}$ (right) mesons for Model 1 in (\ref{case1})
with $K_G=0.002$.
(b) $A_N^D$ for the $D^0$ (left) and $\bar{D^0}$ (right) mesons for Model 2 in (\ref{case2})
with $K_G'=0.0005$. Bars denote the RHIC preliminary data taken from \cite{Liu}.}
\label{fig.case2}
\end{center}
\end{figure}

Fig. 6  shows the results for the contribution to
$A_N^D$ from each term in the twist-3 cross section shown in (\ref{twist3final}) with
$K_G=0.002$ for the model 1 (Fig. 6(a))
and $K_G'=0.0005$ for the model 2 (Fig. 6(b)).      
For both models, one sees that the nonderivative terms accompanying
the hard cross sections $\hat{\sigma}^{O3,O4,N3,N4}$ is negligible
compared to the contributions from $\hat{\sigma}^{O1,O2,N1,N2}$.  
In the left figures of Figs. 6 (a) and (b), 
we have also plotted the contribution from the nonderivative terms
proportional to $\hat{\sigma}^{O1,O2,N1,N2}$, which shows that
the derivative contributions dominate in these terms at large $x_F$ $(>0)$,
while the effect of the nonderivative term becomes important 
at $x_F<0$ and even dominates the asymmetry for the model 1. 
One also sees that $\hat{\sigma}^{O1}$ and $\hat{\sigma}^{O2}$ give
rise to numerically very close asymmetries at the RHIC energy 
for the two models, and 
likewise for $\hat{\sigma}^{N1}$ and $\hat{\sigma}^{N2}$.  
The asymmetries caused by
$\hat{\sigma}^{O1,O2}$ and $\hat{\sigma}^{N1,N2}$ are also similar.

Fig. 7 shows the result for $A_N^D$ for the $D$ and $\bar{D}$ 
mesons including all the contributions
in (\ref{twist3final}) together with the preliminary data by 
the PHENIX collaboration\,\cite{Liu}.  
Because the 
sign of the contribution from $\{O(x,x),O(x,0)\}$
changes between $D$ and $\bar{D}$ as shown in (\ref{twist3final}), 
$\{O(x,x),O(x,0)\}$ and $\{N(x,x),N(x,0)\}$ contribute to the asymmetry
constructively (destructively) for the $D$ ($\bar{D}$) meson, leading to a large (small) $A_N^D$
for $D$ ($\bar{D}$).    
If one reverses the relative sign between $O$ and $N$ from (\ref{OONN}),
the result for the $D$ and $\bar{D}$ mesons will be interchanged.  
The values $K_G=0.002$ and $K'_G=0.0005$ have been chosen such that
the calculated asymmetries 
does not overshoot the data for $A_N$ for the $D$-meson (left figures in Figs. 7(a) and (b))
under the assumption (\ref{OONN}).  
By comparing the results for the models 1 and 2 in Fig. 7, 
one sees that the behavior of the asymmetry at $x_F<0$ depends strongly on
the small-$x$ behavior of the three gluon correlation functions.  
Therefore $A_N^D$ at $x_F<0$ is useful to get constraint on the small-$x$ behavior of the
three-gluon correlation functions.

As we saw in the left figures of Fig. 6, 
we found the relations $\hat{\sigma}^{O1}\simeq \hat{\sigma}^{O2}$
and $\hat{\sigma}^{N1}\simeq \hat{\sigma}^{N2}$.  
This means that the combinations 
$O(x,x)+O(x,0)$ and $N(x,x)-N(x,0)$
can be taken as good effective three-gluon correlation functions
determining $A_N^D$'s at RHIC energies.  
From the left figures in Figs. 7 (a) and (b),
if the ansatz (\ref{case1}) or (\ref{case2}) is a reasonable assumption for the
$x$-dependence of the 
three-gluon correlation functions, 
$K_G=0.002$ or $K'_G=0.0005$ can be taken as a modest upper bound 
corresponding to the assumptions (\ref{case1}) and (\ref{case2}).  
Therefore we may set the upper bound for the combination as 
\beq
|O(x,x)+O(x,0)|\leq 0.004\,x\,G(x),\qquad |N(x,x)-N(x,0)|\leq 0.004\,x\,G(x),
\label{constraint1}
\eeq
and
\beq
|O(x,x)+O(x,0)|\leq 0.001\sqrt{x}\,G(x),\qquad |N(x,x)-N(x,0)|\leq 0.001\sqrt{x}\,G(x),
\label{constraint2}
\eeq
although extraction of the separate constraint on the four functions is not possible.  
\footnote{In \cite{Koike:2010jz}, we extracted a stronger constraint for the upper bound of 
$|O(x,x)+O(x,0)|$ and $|N(x,x)-N(x,0)|$, since we assumed $|A_N^D|<5$ \% in the wider region
of $x_F$, while the data showing $|A_N^D|<5$ \% is only in the region $|x_F|<0.1$\,\cite{Liu}.    
Therefore, at present, a weaker upper bound in (\ref{constraint1}) and (\ref{constraint2}) 
is more appropriate. }
We remark that even though the RHIC data suggests small $A_N^D$ at $|x_F|<0.1$,
it can be much larger at $|x_F|>0.1$ depending on the
behavior of the three-gluon correlation functions in
the large and small $x$ regions as shown in Fig. 7.

\section{Summary}
In this paper we have studied the SSA for the open-charm production
in the $pp$ collision, $p^\uparrow p\to DX$, 
based on the twist-3 mechanism in the collinear factorization.
Since the three-gluon correlation functions in the transversely polarized nucleon
play a dominant role in giving rise to 
SSA for this process, we have derived the corresponding twist-3 single-spin-dependent
cross section in the leading order QCD.
As in the case of our previous study on $ep^\uparrow\to eDX$, 
our result differs from the existing result in the literature.  
We have also derived the master formula which shows that the corresponding twist-3 cross section 
can be obtained from the hard part for the $gg\to c\bar{c}$ scattering in the twist-2 level.
The use of this formula simplifies the actual calculation and is useful
to make the structure of the twist-3 cross section transparent.  
We expect that this master formula is useful for the inclusion of the higher-order corrections to
the cross section.
We also presented a model calculation of the
asymmetry $A_N^D$ in comparison to the preliminary data obtained at RHIC.
We have shown that $A_N^D$ at $x_F<0$ is sensitive to the small-$x$ behavior
of the three-gluon correlation functions, and have given a modest upper limit
on those functions.

\section*{Acknowledgments}
We thank D. Boer, Z.-B. Kang, K. Tanaka, M. Liu, J.-W. Qiu and
F. Yuan for useful discussions, and the authors of Ref. \cite{KKKS08} for providing us with
the Fortran code of their $D$-meson fragmentation function.  
The work of S. Y. is supported by the Grand-in-Aid for Scientific Research
(No. 22.6032) from the Japan Society of Promotion of Science.

\appendix

\section{Ward identity for the initial state interaction}

To derive the cross section for $p^\uparrow p\to DX$, 
one has to analyze
\beq
\int{d^4 k_1\over (2\pi)^4}\int{d^4 k_2\over (2\pi)^4}\,
S_{\mu\nu\lambda}^{abc}(k_1,k_2,x'p',p_c)M_{abc}^{\mu\nu\lambda}(k_1,k_2),
\label{start}
\eeq
where $M_{abc}^{\mu\nu\lambda}(k_1,k_2)$ is defined in (\ref{AAA}) 
and $S_{\mu\nu\lambda}^{abc}(k_1,k_2,x'p',p_c)$ is the
corresponding hard part as shown in Fig. 3.  
As was shown in \cite{BKTY10},  
in order to be able to obtain the cross section from (\ref{twist3}), it is essential 
that $S_{\mu\nu\lambda}^{abc}(k_1,k_2,x'p',p_c)$ satisfies the Ward identities
\beq
k_1^\mu S_{\mu\nu\lambda}^{abc} (k_1,k_2)=0,\qquad
k_2^\nu S_{\mu\nu\lambda}^{abc} (k_1,k_2)=0,\qquad
(k_2-k_1)^\lambda S_{\mu\nu\lambda}^{abc}(k_1,k_2)=0.  
\label{ward}
\eeq  
It is easy to see that the 
hard part for the sum of the FSI diagrams shown in Fig. 4(a) 
satisfies
 (\ref{ward}) due to the on-shell condition for the bared quark-lines.  
However, the hard part for the ISI diagrams in Fig. 4(b) does not satisfy (\ref{ward}).  
This is because the polarization tensor for the gluon line producing the 
SGP (bared gluon line in Fig. 4(b)) is taken to be $-g_{\sigma\tau}$
in the Feynman gauge, which contains the contribution from unphysical polarizations.  
Nevertheless, one can calculate the hard cross section for the ISI from (\ref{twist3}).  
To show this, we define the new hard 
part for the ISI, $\widetilde{S}_{\mu\nu\lambda}^{I,abc}(k_1,k_2,x'p',p_c)$,
which is obtained from the original hard part for ISI, $S_{\mu\nu\lambda}^{I,abc}(k_1,k_2,x'p',p_c)$, by 
replacing the polarization tensor
for the bared gluon-propagator in Fig. 4(b) as
\beq
-g_{\sigma\tau} \rightarrow -g_{\sigma\tau} + {q_\sigma p_\tau + q_\tau p_\sigma
\over q\cdot p},
\label{replace}
\eeq
where $q$ is the momentum carried by the bared gluon-line.  
One can show that in the twist-3 accuracy this $\widetilde{S}_{\mu\nu\lambda}^{I,abc}(k_1,k_2,x'p',p_c)$
satisfies the relation
\beq
&&\int{d^4 k_1\over (2\pi)^4}\int{d^4 k_2\over (2\pi)^4}\,
S_{\mu\nu\lambda}^{I,abc}(k_1,k_2,x'p',p_c)M^{\mu\nu\lambda}_{abc}(k_1,k_2) \nonumber\\
&&\qquad\qquad\qquad =\int{d^4 k_1\over (2\pi)^4}\int{d^4 k_2\over (2\pi)^4}\,
\widetilde{S}_{\mu\nu\lambda}^{I,abc}(k_1,k_2,x'p',p_c)M^{\mu\nu\lambda}_{abc}(k_1,k_2),
\eeq
where $M^{\mu\nu\lambda}_{abc}(k_1,k_2) $ is defined in (\ref{AAA}).
The extra terms in $\widetilde{S}_{\mu\nu\lambda}^{I,abc}(k_1,k_2,x'p',p_c)$
compared to ${S}_{\mu\nu\lambda}^{I,abc}(k_1,k_2,x'p',p_c)$ which occur due to the replacement (\ref{replace})
can be shown to vanish
with the help of the Feynman gauge condition
for the matrix element:
\beq
k_{1\mu} M^{\mu\nu\lambda}_{abc}(k_1,k_2)=0,\quad
k_{2\nu} M^{\mu\nu\lambda}_{abc}(k_1,k_2)=0,
\quad
{(k_2-k_1)}_\lambda M^{\mu\nu\lambda}_{abc}(k_1,k_2)=0.  
\eeq
From the above relation, one can use $\widetilde{S}_{\mu\nu\lambda}^{I,abc}(k_1,k_2,x'p',p_c)$
for the calculation of the twist-3 cross section.  
Since $\widetilde{S}_{\mu\nu\lambda}^{I,abc}(k_1,k_2)$ satisfies
\beq
k_1^\mu \widetilde{S}_{\mu\nu\lambda}^{I,abc} (k_1,k_2)=0,\quad
k_2^\nu \widetilde{S}_{\mu\nu\lambda}^{I,abc} (k_1,k_2)=0,\quad
(k_2-k_1)^\lambda \widetilde{S}_{\mu\nu\lambda}^{I,abc}(k_1,k_2)=0,
\eeq
the resulting twist-3 cross section from the ISI takes the form of
(\ref{twist3}) with $S_{\mu\nu\lambda}^{abc}(k_1,k_2,x'p',p_c)$
replaced by $\widetilde{S}_{\mu\nu\lambda}^{I,abc}(k_1,k_2,x'p',p_c)$.
Finally, one can show the relation
\beq
&&\left.
{\partial
\widetilde{S}_{\mu\nu\lambda}^{I,abc}(k_1,k_2,x'p',p_c)p^{\lambda}\over \partial k_2^{\sigma}}\right|_{k_i=x_ip}
\omega^\mu_{\ \,\alpha}\omega^\nu_{\ \,\beta}\omega^\sigma_{\ \,\gamma}
M^{\alpha\beta\gamma}_{F,abc}(x_1,x_2)\nonumber\\
&&\qquad\qquad\qquad=
\left.
{\partial
S_{\mu\nu\lambda}^{I,abc}(k_1,k_2,x'p',p_c)p^{\lambda}\over \partial k_2^{\sigma}}\right|_{k_i=x_ip}
\omega^\mu_{\ \,\alpha}\omega^\nu_{\ \,\beta}\omega^\sigma_{\ \,\gamma}
M^{\alpha\beta\gamma}_{F,abc}(x_1,x_2), 
\eeq
where the right-hand-side is
calculated with the original hard part for the ISI.  
This way one can obtain the contribution of the three-gluon correlation functions
to the twist-3 cross section 
from (\ref{twist3}).


\end{document}